\batchmode
\makeatletter
\def\input@path{{C:/Users/kenxy/Dropbox/yamada/06.payrise/paper/submission/arXiv/}}
\makeatother
\documentclass[12pt,english]{article}
\usepackage{mathptmx}

\usepackage[T1]{fontenc}
\usepackage[latin9]{inputenc}
\usepackage[svgnames]{xcolor}
\usepackage{colortbl}
\usepackage{babel}
\usepackage{amsmath}
\usepackage{amsthm}
\usepackage{amssymb}
\usepackage{graphicx}
\usepackage[a4paper]{geometry}
\usepackage{setspace}
\usepackage[authoryear]{natbib}
\usepackage[pdftex,pdfusetitle,
 bookmarks=true,bookmarksnumbered=false,bookmarksopen=false,
 breaklinks=true,pdfborder={0 0 0},pdfborderstyle={},backref=false,colorlinks=true]
 {hyperref}
\hypersetup{
 citecolor=DarkBlue,linkcolor=DarkBlue,urlcolor=DarkBlue}

\makeatletter

\providecommand{\tabularnewline}{\\}

\usepackage[svgnames]{xcolor}
\usepackage{bbm}

\ifdefined\showcaptionsetup
 \PassOptionsToPackage{caption=false}{subfig}
\fi
\usepackage{subfig}
\makeatother

\theoremstyle{plain}
\newtheorem{prop}{\protect\propositionname}
\providecommand{\propositionname}{Proposition}

\begin{document}
\title{Decomposing Wage Stagnation:\\
Employment Reallocation, Wage Structure,\\
and Demographics\thanks{I am grateful to seminar participants at Doshisha University and the
Economic and Social Research Institute for helpful comments. I gratefully
acknowledge support from JSPS KAKENHI grant number 26K04947.}}
\author{Ken Yamada\thanks{Kyoto University. \texttt{yamada@econ.kyoto-u.ac.jp}}}
\date{October 2026\\
}
\maketitle
\begin{abstract}
\begin{onehalfspace}
Japan's average log real hourly wages rose from the 1980s to the mid-1990s,
declined through the mid-2010s, and partially recovered thereafter.
This paper decomposes these changes from 1980 to 2024 into four components:
demographic change across worker types, employment reallocation across
job types, changes in relative wage structure across job types, and
unweighted mean wage growth. The first two represent different margins
of adjustment in employment, while the latter two represent different
margins of adjustment in wages. The framework combines a shift\textendash share
decomposition across worker types with an augmented Olley\textendash Pakes
decomposition across job types within worker types. Employment reallocation
reflects changes in employment shares across job types, whereas changes
in relative wage structure reflect changes in average log wages within
job types that alter wage differences across job types. The contributions
of the four components vary across periods. Before 1996, unweighted
mean wage growth and changes in relative wage structure contribute
positively, while demographic change and employment reallocation contribute
negatively. During the 1996\textendash 2014 period, unweighted mean
wage growth and changes in relative wage structure turn negative,
so that all four components are negative. After 2014, the recovery
mainly reflects unweighted mean wage growth. The contributions from
employment reallocation and changes in relative wage structure differ
not only between full-time and part-time jobs but also across establishment
sizes and industries within employment type. The paper uses the same
framework to decompose changes in Japan\textquoteright s gender wage
gap.\bigskip{}
\end{onehalfspace}

\global\long\def\E{\mathbb{E}}%
\end{abstract}
\newpage{}

\section{Introduction}

Advanced economies have experienced wage stagnation alongside changes
in the demographic composition of workers and their allocation across
jobs, as well as changes in wage levels within jobs and wage differences
across jobs. These forces jointly shape aggregate wage growth. While
much of the labor economics literature focuses on average log wages
within groups defined by worker or job attributes, this paper treats
average log real hourly wages for the economy as a whole as the main
object of interest. Average log wages can be expressed as weighted
averages of log wages within groups defined by worker and job attributes.
This representation highlights the roles of both employment shares
and wages across worker and job types. As shown below, changes in
average log wages can be decomposed along four margins: who works,
what kinds of jobs they hold, how wages differ across those jobs,
and how much those jobs pay on average.\footnote{This paper focuses on log wages, as is common in empirical wage analysis.
Changes in log wages are expressed in log points and can be interpreted
approximately as percentage changes. The decomposition itself can
also be applied to wages in levels, in which case changes are expressed
in units of currency.} Viewed through these four margins, average log wages may stagnate
because employment shifts toward lower-wage worker types, employment
within worker types shifts toward lower-wage job types, relative wages
decline for jobs with higher employment shares, or wages fail to rise
on average across job types.

Japan's experience brings these margins into sharp relief. Real wages
have stagnated since the mid-1990s amid an aging workforce, rising
female employment, and expanding part-time employment.\footnote{The growing prevalence of part-time employment took place against
a background of prolonged labor-market adjustment, rising employment
among women and older workers, retirement and continued-employment
practices, and regulatory changes that broadened the scope for the
use of nonregular employment.} An aging workforce and rising female employment relate to changes
in which types of workers are employed, while expanding part-time
employment relates to changes in the kinds of jobs held by workers
of a given type. At the same time, the gender wage gap has narrowed,
wage growth with age has weakened, and the extent of these changes
varies across job types. These wage changes relate to changes in wage
levels within job types and in wage differences across job types for
workers of a given type.

This paper develops a decomposition framework that proceeds in three
steps. The framework first applies a shift\textendash share decomposition
across worker types, then applies an Olley\textendash Pakes decomposition
across job types within each worker type, and finally applies a variant
of the shift\textendash share decomposition to changes over time in
the allocation term in the Olley\textendash Pakes decomposition. I
refer to the combination of the decompositions in the second and third
steps as the augmented Olley\textendash Pakes decomposition. The three-stage
decomposition separates changes in average log wages into four components:
demographic change across worker types, changes in relative employment
shares across job types, changes in relative log wages across job
types, and the unweighted mean of wage growth by job type. The first
two represent different margins of adjustment in employment across
worker types and across job types within worker types. The latter
two represent different margins of adjustment in wages across job
types and within job types for each worker type. Different classes
of labor-market models emphasize different margins of adjustment.
Typically, models of labor supply feature participation decisions
that can shape the composition of employment across worker types \citep{Blundell_MaCurdy_HLE99},
while models of worker assignment and of search and matching focus
on mechanisms that shape the allocation of workers across job types
\citep{Sattinger_JEL93,Mortensen_Pissarides_HLE99}. Models of human
capital and of rent sharing feature mechanisms that can shape the
wage structure across job types as well as wage growth within job
types \citep{Willis_HLE86,Card_Cardoso_Heining_Kline_JoLE18}.

The main contribution of the paper is to provide a unified accounting
framework for measuring the contributions of distinct margins to changes
in average log wages. The framework has four key features. First,
worker types and job types play different roles. Worker types are
defined by demographic characteristics across which employment shares
change, while job types are defined by job and workplace characteristics
across which employment shares and wages vary within worker types.
In the analysis, worker types are defined by sex and age group, while
job types are defined by employment type, establishment size, and
industry. The choice of attributes used to define worker and job types
can be guided by the research question and by the set of characteristics
consistently available in the data. Second, the Olley\textendash Pakes
decomposition is applied across job types within worker types, rather
than across firms in its standard use in productivity analysis. Workers
differ not only in job and workplace characteristics but also in demographic
characteristics. The shift\textendash share decomposition across worker
types separates changes in the demographic composition of workers
from changes in average log wages within worker types. The change
in the allocation term is then decomposed into two components: employment
reallocation and changes in relative wage structure. Employment reallocation
reflects changes in employment shares across job types, whereas changes
in relative wage structure reflect changes in average log wages within
job types that alter wage differences across job types. These two
components can contribute differently to changes in average log wages
and work in opposite directions. Without this distinction, the offsetting
effects of these two margins would be masked. Third, the decomposition
into these four components is invariant to the order in which the
version of the shift\textendash share decomposition used here and
the Olley\textendash Pakes decomposition are applied. This property
links the two decompositions. For a given worker type, changes in
relative wage structure and unweighted mean wage growth together correspond
to changes in average log wages within job types in the shift\textendash share
decomposition, while employment reallocation and changes in relative
wage structure together correspond to the change in the allocation
term in the Olley\textendash Pakes decomposition. Fourth, the framework
can also be applied to study the sources of changes in between-group
inequality, measured by differences in average outcomes across groups
defined by worker characteristics. I use the framework to decompose
changes in Japan\textquoteright s gender wage gap, which is among
the largest in the OECD.\footnote{See OECD, \textquotedblleft Gender wage gap,\textquotedblright{} https://www.oecd.org/en/data/indicators/gender-wage-gap.html
(accessed October 9, 2026).}

Applying the decomposition framework developed in this paper to repeated
cross-sectional data from a large-scale government survey in Japan
over the 45-year span from 1980 to 2024, I show that the contributions
of the four components to changes in average log wages vary markedly
across periods. Before 1996, unweighted mean wage growth and changes
in relative wage structure contribute positively, more than offsetting
negative contributions from demographic change and employment reallocation.
During the 1996\textendash 2014 period, unweighted mean wage growth
and changes in relative wage structure turn negative, so that all
four components are negative. After 2014, the recovery mainly reflects
renewed unweighted mean wage growth. The substantial decline in the
employment share of prime-age men, who have the highest wage levels,
is a major factor behind the negative contribution from demographic
change. Employment reallocation lowers average log wages before 2014,
while changes in relative wage structure lower them after 1996. Neither
pattern is limited to the distinction between full-time and part-time
jobs. The contributions of both components also differ across establishment
sizes and industries within employment type. Taken together, the results
show that Japan\textquoteright s wage stagnation reflects the joint
evolution of worker composition, employment allocation, wage structure,
and wage growth.

I further show that the contributions of the four components to changes
in the male\textendash female wage gap vary markedly across periods.
Before 1996, employment reallocation and changes in relative wage
structure contribute positively, largely offsetting the negative contribution
from unweighted mean wage growth. During the 1996\textendash 2014
period, employment reallocation and changes in relative wage structure
turn negative, while unweighted mean wage growth remains negative
and demographic change remains positive. After 2014, the narrowing
mainly reflects negative contributions from unweighted mean wage growth
and employment reallocation, while changes in relative wage structure
contribute positively and demographic change is slightly negative.
Much of the narrowing after 1996 is concentrated among workers aged
30 to 54. The narrowing of the gender wage gap mainly reflects unweighted
mean wage growth rather than employment reallocation across job types,
even as the share of part-time employment increases more among women
than among men within each age group.

The remainder of the paper is organized as follows. Section 2 summarizes
the related literature. Section 3 describes the data and descriptive
evidence. Section 4 presents the decomposition framework and statistical
inference. Section 5 reports the empirical results. Section 6 concludes.

\section{Related Literature\label{sec: literature}}

This paper relates to two strands of the literature. The first strand
studies nonstandard employment, labor market dualization, and wage-setting
practices. Nonstandard employment arrangements, including part-time,
temporary, and contract work, have long been studied as forms of employment
that differ from standard employment in terms of wages, security,
training, and career prospects \citep{Kalleberg_ARS00,OECD_bk14,ILO_bk16}.
In economics, this discussion is closely related to the literature
on labor market dualism, which emphasizes institutional divisions
between more protected and less protected segments of the labor market
\citep{Bentolila_Dolado_EP94,Boeri_HLE11}. International evidence
also documents wage differences between standard and nonstandard jobs
\citep{Kahn_IR16}. Prior work in Japan has emphasized the distinction
between regular and nonregular employment, the role of long-term employment
and job security, and the evolution of seniority-based wage differentials
\citep{Hamaaki_etal_ILRR12,Asano_Ito_Kawaguchi_SJPE13,Yamada_Kawaguchi_JEI15,Kambayashi_Kato_ILRR17}.
This paper builds on this literature but takes a broader view. Employment
type is treated as one dimension of job type, alongside establishment
size and industry. The analysis examines how employment reallocation
and changes in relative wage structure across this broader set of
job types contribute to Japan's long-run wage stagnation.

The second strand is the literature on wage decomposition. Much of
this literature relies on regression-based approaches, most notably
the Oaxaca\textendash Blinder decomposition, which separates wage
differences into components associated with differences in worker
characteristics and differences in wage structures \citep{Oaxaca_IER73,Blinder_JHR73}.
The objective here is to provide an accounting decomposition of changes
in average log real hourly wages, with an application to changes in
the gender wage gap. To this end, two approaches are relevant. The
shift\textendash share decomposition separates aggregate changes into
between- and within-group components \citep{Bennet_JRSS20,Fabricant_bk42,Kitagawa_JASA55},
while the Olley\textendash Pakes decomposition separates an unweighted
mean outcome from an allocation term \citep{Olley_Pakes_EM96}. The
dynamic Olley\textendash Pakes decomposition of \citet{Melitz_Polanec_RJE15}
extends this decomposition to changes in aggregate productivity when
production units enter and exit. Shift\textendash share decompositions,
including Bennet-type decompositions, have been used to study the
sources of changes in wages and employment \citep{Katz_Murphy_QJE92,Berman_Bound_Griliches_QJE94,Daly_Hobijn_AERPP17}.
The Olley\textendash Pakes and Melitz\textendash Polanec decompositions
were developed for the analysis of aggregate productivity but have
also been applied to study the sources of changes in wages and labor
shares \citep{Adamopoulou_Bobbio_DePhilippis_Giorgi_IZA,Kehrig_Vincent_QJE21}.\footnote{\citet{Kehrig_Vincent_QJE21} and the working-paper version of \citet{Adamopoulou_Bobbio_DePhilippis_Giorgi_IZA}
decompose the change in the allocation term in the Olley\textendash Pakes
decomposition into components associated with changes in outcomes,
changes in shares, and an interaction term.} I develop a decomposition framework that combines the shift\textendash share
decomposition across worker types with the augmented Olley\textendash Pakes
decomposition across job types within worker types. In the augmented
Olley\textendash Pakes decomposition, the change in the allocation
term is expressed exactly as the sum of two components that represent
distinct margins of adjustment in employment and wages, without a
separate interaction term. This framework exhibits an order-invariance
property: for a fixed set of groups, the decomposition of the change
in a weighted average is invariant to the order in which the version
of the shift\textendash share decomposition used here and the Olley\textendash Pakes
decomposition are applied. When applied across job types, standard
shift\textendash share decompositions do not distinguish between changes
in relative wage structure and unweighted mean wage growth, while
the standard Olley\textendash Pakes decomposition does not distinguish
between changes in relative wage structure and employment reallocation.
The framework ultimately separates changes in average log wages into
demographic change across worker types, employment reallocation across
job types, changes in relative wage structure across job types, and
the unweighted mean of wage growth by job type.

\section{Data and Descriptive Evidence\label{sec: data}}

The analysis uses repeated cross-sectional data from the Basic Survey
on Wage Structure (BSWS), a large-scale government survey administered
by Japan's Ministry of Health, Labour and Welfare, covering the period
from 1980 to 2024. The BSWS covers employees in establishments with
five or more regularly employed workers in nearly all industries except
agriculture, forestry, fisheries, and government (except elsewhere
classified). The data used in the analysis are available for 1980,
1985, and annually from 1989 onward. The BSWS provides detailed information
on wages, hours worked, employment type, establishment size, industry,
and prefecture. Hourly wages are calculated as monthly earnings plus
one-twelfth of annual bonuses, divided by monthly hours worked. Wages
are deflated by the consumer price index and expressed in yen at 2020
prices. The sample consists of employees aged 16 to 74 and excludes
those with extreme values for monthly hours worked or hourly wages.\footnote{The sample excludes employees whose monthly hours worked are below
10 or above 350, or whose hourly wages are below one-half of the statutory
minimum wage or at or above 50,000 yen at 2020 prices.} The number of employees in each survey year ranges from 1.1 million
to 1.4 million. Each employee is weighted using the survey sampling
weight.

Worker types are defined by sex and age group: men and women aged
16 to 29, 30 to 54, and 55 to 74. Job types are defined by employment
type, establishment size, and industry. Employment type distinguishes
full-time and part-time jobs, corresponding to the survey distinction
between general workers and part-time workers that is consistently
available over the 1980\textendash 2024 period. Establishment size
distinguishes establishments with fewer than 30 regularly employed
workers from those with 30 or more. Industries are grouped into five
broad categories: manufacturing; infrastructure, which combines mining,
construction, electricity, and transportation; wholesale; finance
and real estate; and services. This classification by multiple job
attributes makes it possible to examine whether employment reallocation
and changes in relative wage structure are limited to the distinction
between full-time and part-time jobs or also extend across industries
and establishment sizes within employment type. The classification
is consistent across survey years and is intentionally kept broad
to maintain balanced support across worker types and job types over
time. Before implementing the decomposition, I verify that every cell
formed by worker type and job type is observed in every sample year.

\begin{figure}[h]
\caption{Hourly wages and employment shares by worker type\label{fig: wage=000026share}}

\begin{centering}
\subfloat[Hourly wages]%
{
\centering{}\includegraphics[scale=0.58]{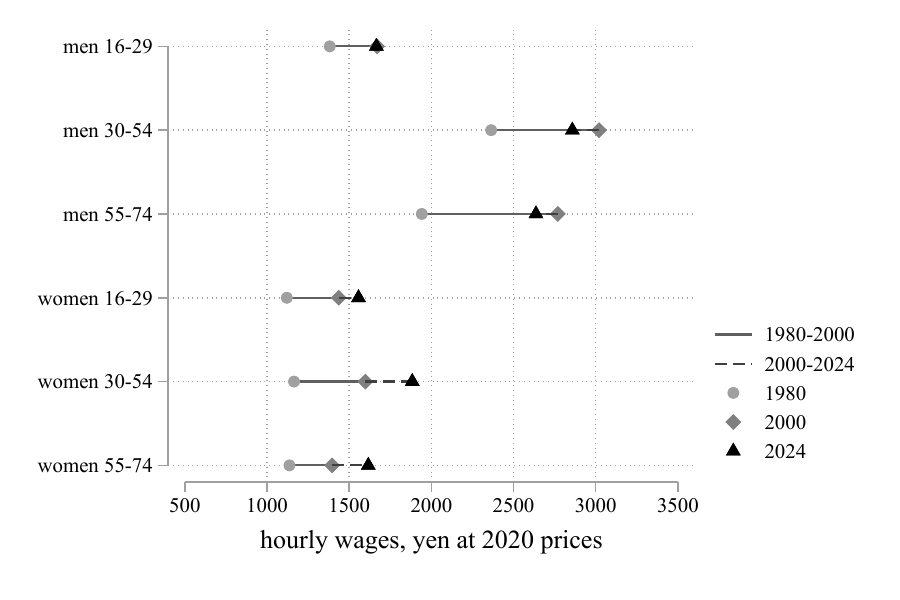}}%
\subfloat[Employment shares]%
{
\centering{}\includegraphics[scale=0.58]{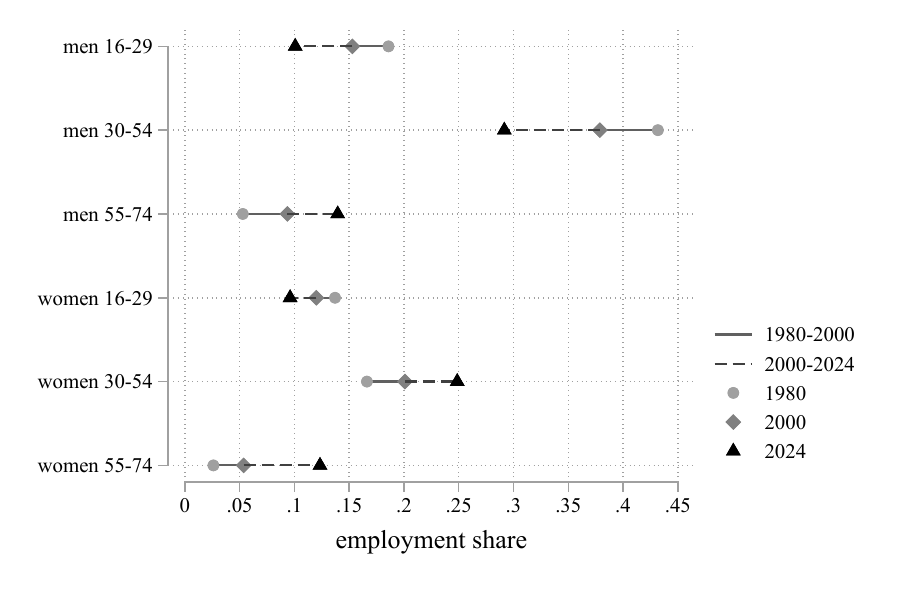}}
\par\end{centering}
{\scriptsize\textit{Notes}}{\scriptsize : Hourly wages are measured
in yen and deflated to 2020 prices. Employment shares are measured
as shares of total employment. The line segments connect 1980 to 2000
and 2000 to 2024, respectively.}{\scriptsize\par}
\end{figure}

Figure \ref{fig: wage=000026share} summarizes the key descriptive
patterns of the levels and changes in hourly wages and employment
shares by worker type. The line segments connect 1980 to 2000 and
2000 to 2024, respectively, to summarize changes over the earlier
and later periods. Panel (a) shows that real hourly wages differ substantially
across worker types and that wages do not increase monotonically over
the 1980\textendash 2024 period. For several worker types, wages increase
between 1980 and 2000 but decline or stagnate after 2000, especially
among men. The wage differences between men and women narrow substantially
over the period for the younger and prime-age groups. Panel (b) shows
more monotonic changes in employment shares. Employment shifts away
from prime-age men and toward women and older workers. These patterns
imply that changes in average log wages depend not only on wage changes
within worker types, but also on how employment is distributed across
worker types. Figures \ref{fig: wage=000026emp1_men} to \ref{fig: wage=000026emp2_women}
in the appendix provide the corresponding descriptive evidence by
job type. Figures \ref{fig: wage=000026emp1_men} and \ref{fig: wage=000026emp1_women}
show variation by employment type and industry, while Figures \ref{fig: wage=000026emp2_men}
and \ref{fig: wage=000026emp2_women} show variation by employment
type and establishment size. Together, they show that wage levels
differ substantially not only between full-time and part-time jobs,
but also across industries and establishment sizes within employment
type. Employment shares also vary across job types and change over
time, with different patterns for men and women. These patterns support
the distinction between worker types and job types and provide a basis
for examining employment reallocation and changes in relative wage
structure across job types within worker types.

\section{Decomposition Framework\label{sec: framework}}

This section presents the accounting framework for decomposing changes
in average log wages. The framework first separates the part attributable
to changes in the demographic composition of workers across worker
types from the part attributable to changes in average log wages within
worker types. It then decomposes the part attributable to changes
in average log wages within worker types into three components: changes
in relative employment shares across job types, changes in relative
log wages across job types, and the unweighted mean of changes in
log wages by job type. Together, these steps separate demographic
change, employment reallocation, changes in relative wage structure,
and unweighted mean wage growth.

\subsection{Decomposition into four components\label{subsec: decomp}}

Let $w_{ghit}$ denote the logarithm of the real hourly wages of individual
$i$ in worker type $g$ (such as sex and age), job type $h$ (such
as employment type, establishment size, and industry), and year $t$.
Average log wages can be written as the following weighted averages:
\[
w_{t}=\sum_{g}s_{gt}w_{gt},\quad w_{gt}=\sum_{h}s_{ght}w_{ght},\quad w_{ght}=\sum_{i}s_{ghit}w_{ghit}.
\]
The employment shares are given by
\[
s_{gt}=\frac{n_{gt}}{\sum_{g}n_{gt}},\quad s_{ght}=\frac{n_{ght}}{\sum_{h}n_{ght}},\quad s_{ghit}=\frac{n_{ghit}}{\sum_{i}n_{ghit}},
\]
where $n_{ghit}$ denotes the sampling weight assigned to individual
$i$, $n_{ght}=\sum_{i}n_{ghit}$, and $n_{gt}=\sum_{h}n_{ght}$.

I first decompose the change in average log wages by applying a shift\textendash share
decomposition across worker types. Specifically, I use the Kitagawa
decomposition, which expresses this change as exactly two components
using midpoint averages \citep{Kitagawa_JASA55}. This step separates
changes in average log wages that arise from changes in the demographic
composition of workers across worker types from those that arise from
changes in average log wages within each worker type. The change in
average log wages can be decomposed as
\[
\Delta w_{t}=\sum_{g}\check{w}_{g}\Delta s_{gt}+\sum_{g}\check{s}_{g}\Delta w_{gt},
\]
where the midpoint averages are expressed as
\[
\check{w}_{g}=\frac{w_{gt}+w_{g,t-1}}{2},\quad\check{s}_{g}=\frac{s_{gt}+s_{g,t-1}}{2}.
\]
The first term is attributable to changes in employment shares across
worker types, and the second term is attributable to changes in average
log wages within worker types.

I further decompose the change in average log wages within worker
types by applying the standard Olley\textendash Pakes decomposition
across job types \citep*{Olley_Pakes_EM96}. This step separates changes
in average log wages within worker types that arise from changes in
the unweighted average of log wages across job types from those that
arise from changes in the covariation between employment shares and
average log wages across job types. Let $H$ denote the number of
job types. Average log wages within worker types can be decomposed
as
\[
w_{gt}=\overline{w}_{gt}+\sum_{h}\widetilde{s}_{ght}\widetilde{w}_{ght},
\]
where the unweighted averages are expressed as
\[
\overline{w}_{gt}=\frac{1}{H}\sum_{h}w_{ght},\quad\overline{s}_{gt}=\frac{1}{H}\sum_{h}s_{ght}=\frac{1}{H},
\]
and the deviations from these unweighted averages are expressed as
\[
\widetilde{w}_{ght}=w_{ght}-\overline{w}_{gt},\quad\widetilde{s}_{ght}=s_{ght}-\overline{s}_{gt}.
\]
The first term is the unweighted average of log wages across job types
within each worker type. Its change is therefore the unweighted average
of changes in log wages by job type.

The change in average log wages can then be decomposed as
\[
\Delta w_{t}=\sum_{g}\check{w}_{g}\Delta s_{gt}+\sum_{g}\check{s}_{g}\Delta\overline{w}_{gt}+\sum_{g}\check{s}_{g}\Delta\sum_{h}\widetilde{s}_{ght}\widetilde{w}_{ght}.
\]
Unlike in the Melitz\textendash Polanec decomposition with entry and
exit, the set of job types is kept fixed by construction. I then augment
the standard Olley\textendash Pakes decomposition by decomposing the
change in the allocation term within each worker type into a part
attributable to changes in relative employment shares across job types
and a part attributable to changes in relative log wages across job
types. I do so by applying a variant of the Kitagawa decomposition.
Specifically, the change in the allocation term can be decomposed
as
\[
\Delta\sum_{h}\widetilde{s}_{ght}\widetilde{w}_{ght}=\sum_{h}\check{\widetilde{w}}_{gh}\Delta\widetilde{s}_{ght}+\sum_{h}\check{\widetilde{s}}_{gh}\Delta\widetilde{w}_{ght},
\]
where $\check{\tilde{w}}_{gh}=(\tilde{w}_{ght}+\tilde{w}_{gh,t-1})/2$
and $\check{\tilde{s}}_{gh}=(\tilde{s}_{ght}+\tilde{s}_{gh,t-1})/2$.
This decomposition shows that the change in the allocation term can
arise not only from changes in relative employment shares across job
types but also from changes in relative log wages across job types.
Separating these two margins makes it possible to distinguish employment
reallocation from changes in relative wage structure across job types.

The following decomposition identity summarizes the accounting framework.
For fixed sets of worker types and job types, and using midpoint weights
for changes in shares and wages, the change in average log wages can
be written as the sum of four components:
\begin{equation}
\Delta w_{t}=\underset{\text{demographic change}}{\underbrace{\sum_{g}\check{w}_{g}\Delta s_{gt}}}+\underset{\text{relative-share change}}{\underbrace{\sum_{g}\check{s}_{g}\sum_{h}\check{\widetilde{w}}_{gh}\Delta\widetilde{s}_{ght}}}+\underset{\text{relative-wage change}}{\underbrace{\sum_{g}\check{s}_{g}\sum_{h}\check{\widetilde{s}}_{gh}\Delta\widetilde{w}_{ght}}}+\underset{\text{cell-wage change}}{\underbrace{\sum_{g}\check{s}_{g}\Delta\overline{w}_{gt}}}.\label{eq: formula}
\end{equation}
Equation \eqref{eq: formula} decomposes the change in average log
wages into four components. The first term captures changes in the
demographic composition of workers. The sign of this component depends
on whether employment shifts toward worker types with higher or lower
wage levels.\footnote{The demographic component in equation \eqref{eq: formula} does not
depend on the unit in which wages are measured because employment
shares across worker types sum to one. When the demographic component
is decomposed by worker type, however, each term $\check{w}_{g}\Delta s_{gt}$
depends on the unit in which wages are measured. For the decomposition
by worker type, I subtract {\scriptsize$\check{w}=(w_{t}+w_{t-1})/2$}.
This normalization leaves the demographic component unchanged.} A shift toward higher-wage worker types raises average log wages,
while a shift toward lower-wage worker types lowers them. The second
term captures changes in relative employment shares across job types
within each worker type, holding relative log wages fixed at their
midpoint averages. The third term captures changes in relative log
wages across job types within each worker type, holding relative employment
shares fixed at their midpoint averages. The fourth term captures
the unweighted mean of changes in log wages by job type within each
worker type. I refer to this term as the cell-wage component because
it is formed from changes in log wages in cells defined by worker
type and job type and does not depend on employment shares across
job types. The second through fourth terms are aggregated across worker
types using the midpoint averages of employment shares.

In the decomposition above, I decompose the change in average log
wages within worker types into three components by first applying
the Olley\textendash Pakes decomposition and then a variant of the
Kitagawa decomposition. The same three components are obtained if
the Kitagawa decomposition is applied first and the Olley\textendash Pakes
decomposition is applied second.
\begin{prop}
For a fixed set of groups, the two-stage decomposition of the change
in a weighted average into three components is invariant to the order
in which the Kitagawa decomposition and the Olley\textendash Pakes
decomposition are applied.
\end{prop}
\begin{proof}
Suppose that the Kitagawa decomposition is applied first. Within worker
type $g$, the change in average log wages can be written as
\[
\Delta w_{gt}=\sum_{h}\check{w}_{gh}\Delta s_{ght}+\sum_{h}\check{s}_{gh}\Delta w_{ght},
\]
where $\check{w}_{gh}=(w_{ght}+w_{gh,t-1})/2$ and $\check{s}_{gh}=(s_{ght}+s_{gh,t-1})/2$.
Employment shares sum to one across job types, and the set of job
types is fixed; that is, $\sum_{h}\Delta s_{ght}=0$ and $\Delta\widetilde{s}_{ght}=\Delta s_{ght}$.
The first term can thus be rewritten as the relative-share component.
\[
\sum_{h}\check{w}_{gh}\Delta s_{ght}=\sum_{h}\check{\widetilde{w}}_{gh}\Delta\widetilde{s}_{ght}.
\]
The Olley\textendash Pakes decomposition is then applied to the second
term in the Kitagawa decomposition. This term can be decomposed as
\[
\sum_{h}\check{s}_{gh}\Delta w_{ght}=\Delta\overline{w}_{gt}+\sum_{h}\check{\widetilde{s}}_{gh}\Delta\widetilde{w}_{ght}.
\]
The first and second terms correspond to the cell-wage component and
the relative-wage component, respectively.
\end{proof}

The order-invariance property extends to the three-stage decomposition:
the same four components are obtained whether the Olley\textendash Pakes
decomposition across job types within worker types is applied in the
first, second, or third step. The proposition clarifies the interpretation
of the three components of changes in average log wages within worker
types: the relative-share, relative-wage, and cell-wage components.
For a given worker type, the relative-wage and cell-wage components
together correspond to changes in average log wages within job types
in the Kitagawa decomposition across job types, while the relative-share
and relative-wage components together correspond to the change in
the allocation term in the Olley\textendash Pakes decomposition across
job types.

Three additional features of the framework are worth noting. First,
the framework distinguishes worker types from job types. Conventional
shift\textendash share decompositions typically use a single set of
groups to separate the part attributable to changes in employment
shares across groups from the part attributable to changes in log
wages within groups. If worker and job types were combined into a
single set of groups, changes in employment shares would not be separated
into demographic change across worker types and employment reallocation
across job types within worker types. Second, the framework can be
readily applied to examine the sources of changes in between-group
inequality, measured by differences in average outcomes across groups
defined by worker characteristics. In Section \ref{subsec: decomp_gap},
I apply the decomposition framework to changes in the gender wage
gap. Third, the decomposition requires only data on employment shares
and average log wages for each cell formed by worker types and job
types defined consistently across the periods being compared. It therefore
does not require longitudinal data and can be implemented using repeated
cross-sectional data.

\subsection{Inference for the decomposition components\label{subsec: inference}}

For statistical inference, I express the four decomposition components
as functions of two parameters for each cell formed by worker type
and job type in each year, both of which are population moments. \citet{Hyytinen_Ilmakunnas_Maliranta_JRSSA16}
similarly formulate the Olley\textendash Pakes decomposition in terms
of moment conditions and use the generalized method of moments for
inference. In their formulation, the two components of the Olley\textendash Pakes
decomposition correspond directly to two parameters in each period,
whereas here the four decomposition components are functions of the
parameters defined for each cell in each period. In both formulations,
the parameters are just identified. For each year $t$, define the
two corresponding sample moments:
\[
\widehat{\pi}_{ght}=\frac{n_{ght}}{\sum_{g}n_{gt}}=s_{gt}s_{ght},\quad\widehat{\mu}_{ght}=\frac{\sum_{i}w_{ghit}n_{ghit}}{\sum_{g}n_{gt}}=s_{gt}s_{ght}w_{ght}.
\]
The decomposition components are functions of employment shares and
average log wages, which are obtained from these moments as
\[
s_{gt}=\sum_{h}\widehat{\pi}_{ght},\quad s_{ght}=\frac{\widehat{\pi}_{ght}}{s_{gt}},\quad w_{ght}=\frac{\widehat{\mu}_{ght}}{\widehat{\pi}_{ght}}.
\]
Let $\hat{\boldsymbol{\theta}}_{t}$ collect $\hat{\pi}_{ght}$ and
$\hat{\mu}_{ght}$ over all worker types $g$ and job types $h$.
For a comparison between years $t_{1}$ and $t_{2}$, define $\hat{\boldsymbol{\theta}}=(\hat{\boldsymbol{\theta}}^{\prime}_{t_{1}},\hat{\boldsymbol{\theta}}^{\prime}_{t_{2}})^{\prime}$.
Let $\boldsymbol{\theta}$ denote the population counterpart of $\hat{\boldsymbol{\theta}}$.
The four decomposition components can then be written as
\[
G\left(\boldsymbol{\theta}\right)=\left(\sum_{g}\check{w}_{g}\left(s_{gt_{2}}-s_{gt_{1}}\right),\sum_{g}\check{s}_{g}\sum_{h}\check{\widetilde{w}}_{gh}\left(\widetilde{s}_{ght_{2}}-\widetilde{s}_{ght_{1}}\right),\sum_{g}\check{s}_{g}\sum_{h}\check{\widetilde{s}}_{gh}\left(\widetilde{w}_{ght_{2}}-\widetilde{w}_{ght_{1}}\right),\sum_{g}\check{s}_{g}\left(\overline{w}_{gt_{2}}-\overline{w}_{gt_{1}}\right)\right)^{\prime}
\]
where the midpoint quantities are defined over $t_{1}$ and $t_{2}$.
The four elements of $G(\boldsymbol{\theta})$ correspond, respectively,
to demographic change, relative-share change, relative-wage change,
and cell-wage change in equation \eqref{eq: formula}. The covariance
matrix of the four decomposition components is estimated by the delta
method:
\[
\widehat{\mathbb{V}}\left[G\left(\widehat{\boldsymbol{\theta}}\right)\right]=\left.\frac{\partial G\left(\boldsymbol{\theta}\right)}{\partial\boldsymbol{\theta}^{\prime}}\right|_{\boldsymbol{\theta}=\widehat{\boldsymbol{\theta}}}\left(\begin{array}{cc}
\widehat{\mathbb{V}}\left(\widehat{\boldsymbol{\theta}}_{t_{1}}\right) & 0\\
0 & \widehat{\mathbb{V}}\left(\widehat{\boldsymbol{\theta}}_{t_{2}}\right)
\end{array}\right)\left.\frac{\partial G\left(\boldsymbol{\theta}\right)^{\prime}}{\partial\boldsymbol{\theta}}\right|_{\boldsymbol{\theta}=\widehat{\boldsymbol{\theta}}}.
\]
For comparisons across survey years, the covariance matrices of $\hat{\boldsymbol{\theta}}_{t_{1}}$
and $\hat{\boldsymbol{\theta}}_{t_{2}}$ are estimated separately.
Because establishments are sampled anew each year, samples from different
survey years are treated as independent and the covariance between
$\hat{\boldsymbol{\theta}}_{t_{1}}$ and $\hat{\boldsymbol{\theta}}_{t_{2}}$
is set to zero. For each year, the covariance matrix of $\hat{\theta}_{t}$
is estimated jointly across all weighted sample moments, with clustering
at the establishment level, thereby allowing for dependence among
workers in the same establishment. This procedure is used to compute
the standard errors reported in Tables \ref{tab: decomp_type1} to
\ref{tab: decomp_gap} and \ref{tab: decomp_type2}. For the decomposition
by worker type, the decomposition components and their sum are also
written as functions of $\theta$, and their standard errors are obtained
by the delta method.

\section{Results\label{sec: results}}

Section \ref{subsec: decomp_main} reports the decomposition results
for changes in average log wages over time and across worker types.
Section \ref{subsec: relative_emp=000026wage} examines the sources
of employment reallocation and changes in relative wage structure
across job types. Section \ref{subsec: decomp_gap} reports the decomposition
results for changes in the gender wage gap over time and across age
groups.

\subsection{Decomposition of changes in average log wages\label{subsec: decomp_main}}

The main decomposition results are presented in two parts. I first
trace the evolution of average log hourly wages and the four decomposition
components over the 45-year span from 1980 to 2024. I then summarize
the decomposition by worker type for three subperiods: 1980\textendash 1996,
1996\textendash 2014, and 2014\textendash 2024. The years 1996 and
2014 correspond to the peak and trough of average log wages. These
three periods are chosen to follow the path of average log wages rather
than to correspond to formal business-cycle dates. Average log wages
continued to rise for several years after the collapse of the asset-price
bubble, although their pace of increase had already slowed. The three
periods nevertheless broadly correspond to Japan's transition from
the stable-growth and bubble-economy years, through the prolonged
post-bubble stagnation, to the recovery period after the mid-2010s.

\subsubsection{Decomposition over time\label{subsec: decomp_time}}

\begin{figure}[h]
\caption{Decomposition of changes in average log real hourly wages\label{fig: decomp_main}}

\begin{centering}
\includegraphics[scale=0.77]{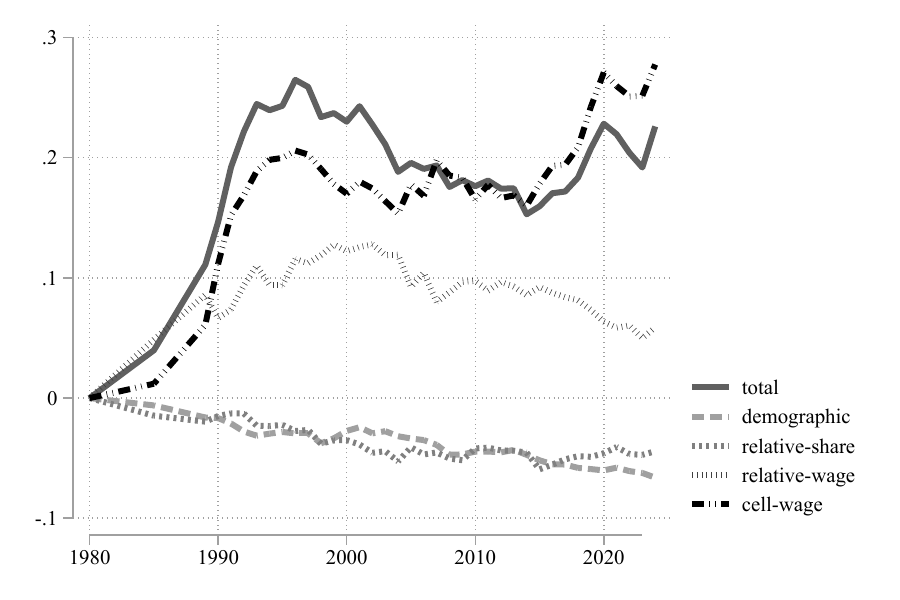}
\par\end{centering}
{\scriptsize\textit{Notes}}{\scriptsize : All series are normalized
to zero in 1980. For each year, the four components decompose the
change in average log wages relative to 1980. The component labels
correspond to the four terms in equation \eqref{eq: formula}: \textquotedblleft demographic\textquotedblright{}
denotes demographic change, \textquotedblleft relative-share\textquotedblright{}
denotes relative-share change, \textquotedblleft relative-wage\textquotedblright{}
denotes relative-wage change, and \textquotedblleft cell-wage\textquotedblright{}
denotes cell-wage change.}{\scriptsize\par}
\end{figure}

Figure \ref{fig: decomp_main} plots the change in average log hourly
wages relative to 1980 and the corresponding four decomposition components.
Average log wages rise rapidly until the mid-1990s, decline through
the mid-2010s, and partially recover thereafter. The contributions
of the four components vary across periods. The cell-wage component
rises strongly before the mid-1990s and after the mid-2010s but declines
between the mid-1990s and the mid-2010s. The demographic and relative-share
components are persistently negative, although the relative-share
component becomes less negative after the mid-2010s. The relative-wage
component rises until around 2000 and declines thereafter, although
it remains above zero. The figure shows that wage stagnation involves
a decline in the cell-wage component between the mid-1990s and the
mid-2010s, persistently negative demographic and relative-share components,
and a decline in the relative-wage component after around 2000. By
2024, the demographic and relative-share components offset about half
of the increase in average log wages relative to 1980. This illustrates
the quantitative importance of demographic change across worker types
and employment reallocation across job types. Figures \ref{fig: decomp_region}
to \ref{fig: decomp_nosize} in the appendix show that the main patterns
are similar to those in Figure \ref{fig: decomp_main} when job types
are classified in alternative ways. Figure \ref{fig: decomposition_nobonus}
in the appendix shows that excluding annual bonuses does not change
the main patterns of the decomposition results, although the decline
in average log wages after the mid-1990s is smaller.

\subsubsection{Decomposition by worker type\label{subsec: decomp_type}}

Table \ref{tab: decomp_type1} reports the decomposition results by
worker type for each period. The sample period is divided into three
phases: wage growth before the mid-1990s, wage decline from 1996 to
2014, and partial recovery after 2014. Between 1980 and 1996, average
log hourly wages increase by 26.5 log points. Unweighted mean wage
growth accounts for most of this increase. The decomposition of the
change in the allocation term shows that its two components work in
opposite directions in this period. The relative-share component is
negative, whereas the relative-wage component is positive. Because
the latter component is larger in absolute value than the former,
the two components sum to a positive value. Without this further decomposition,
the positive value of the allocation term would obscure the fact that
employment reallocation across job types and changes in relative wage
structure work in opposite directions. Demographic change is also
negative, partly offsetting the positive contributions from unweighted
mean wage growth and changes in relative wage structure. Between 1996
and 2014, average log wages fall by 11.2 log points, and all four
components are negative. After 2014, average log wages increase again
by 7.3 log points. The post-2014 increase is mainly accounted for
by the return of positive unweighted mean wage growth. The relative-share
component also becomes slightly positive, while the demographic and
relative-wage components remain negative. The positive estimate of
the relative-share component after 2014 is less precise than the negative
estimates for the earlier periods. The pattern of contributions differs
markedly across the three periods: before 1996, positive contributions
from unweighted mean wage growth and changes in relative wage structure
more than offset negative contributions from demographic change and
employment reallocation; between 1996 and 2014, all four components
are negative; and after 2014, the recovery mainly reflects renewed
unweighted mean wage growth.

\begin{table}[h]
\caption{Contributions by worker type to changes in average log real hourly
wages\label{tab: decomp_type1}}

\begin{centering}
\begin{tabular}{lr@{\extracolsep{0pt}.}lr@{\extracolsep{0pt}.}lr@{\extracolsep{0pt}.}lr@{\extracolsep{0pt}.}lr@{\extracolsep{0pt}.}l}
\hline 
 &
\multicolumn{2}{c}{Total} &
\multicolumn{2}{c}{Demographic} &
\multicolumn{2}{c}{Relative-share} &
\multicolumn{2}{c}{Relative-wage} &
\multicolumn{2}{c}{Cell-wage}\tabularnewline
\cline{2-11}
 &
\multicolumn{10}{c}{1980\textendash 1996}\tabularnewline
Men, 16\textendash 29 &
4&41 (0.08) &
0&37 (0.03) &
\textendash 0&59 (0.04) &
1&00 (0.33) &
3&64 (0.35)\tabularnewline
Men, 30\textendash 54 &
9&84 (0.24) &
\textendash 1&98 (0.10) &
\textendash 0&39 (0.09) &
7&23 (0.88) &
4&99 (0.89)\tabularnewline
Men, 55\textendash 74 &
3&26 (0.07) &
0&46 (0.02) &
\textendash 0&17 (0.02) &
1&29 (0.13) &
1&67 (0.14)\tabularnewline
Women, 16\textendash 29 &
4&20 (0.09) &
0&21 (0.06) &
\textendash 0&54 (0.04) &
0&93 (0.19) &
3&60 (0.20)\tabularnewline
Women, 30\textendash 54 &
4&86 (0.16) &
\textendash 0&89 (0.08) &
\textendash 0&68 (0.09) &
0&98 (0.11) &
5&46 (0.14)\tabularnewline
Women, 55\textendash 74 &
\textendash 0&10 (0.06) &
\textendash 1&10 (0.05) &
\textendash 0&37 (0.02) &
0&15 (0.04) &
1&22 (0.05)\tabularnewline
Total &
26&48 (0.48) &
\textendash 2&94 (0.21) &
\textendash 2&74 (0.20) &
11&56 (1.10) &
20&59 (1.14)\tabularnewline
\cline{2-11}
 &
\multicolumn{10}{c}{1996\textendash 2014}\tabularnewline
Men, 16\textendash 29 &
\textendash 0&63 (0.08) &
1&03 (0.03) &
\textendash 0&89 (0.06) &
\textendash 0&03 (0.09) &
\textendash 0&74 (0.11)\tabularnewline
Men, 30\textendash 54 &
\textendash 7&31 (0.29) &
\textendash 1&36 (0.10) &
\textendash 0&61 (0.10) &
\textendash 0&68 (0.39) &
\textendash 4&66 (0.42)\tabularnewline
Men, 55\textendash 74 &
\textendash 1&42 (0.10) &
0&54 (0.03) &
\textendash 0&79 (0.05) &
\textendash 0&66 (0.10) &
\textendash 0&52 (0.12)\tabularnewline
Women, 16\textendash 29 &
0&28 (0.08) &
1&24 (0.05) &
\textendash 0&72 (0.05) &
\textendash 0&24 (0.07) &
0&00 (0.08)\tabularnewline
Women, 30\textendash 54 &
\textendash 0&36 (0.17) &
\textendash 1&48 (0.07) &
0&45 (0.11) &
\textendash 0&28 (0.09) &
0&94 (0.13)\tabularnewline
Women, 55\textendash 74 &
\textendash 1&74 (0.09) &
\textendash 1&73 (0.06) &
\textendash 0&37 (0.04) &
0&18 (0.05) &
0&18 (0.07)\tabularnewline
Total &
\textendash 11&18 (0.59) &
\textendash 1&76 (0.21) &
\textendash 2&94 (0.27) &
\textendash 1&70 (0.48) &
\textendash 4&79 (0.59)\tabularnewline
\cline{2-11}
 &
\multicolumn{10}{c}{2014\textendash 2024}\tabularnewline
Men, 16\textendash 29 &
0&91 (0.08) &
0&14 (0.03) &
\textendash 0&14 (0.05) &
\textendash 0&13 (0.06) &
1&04 (0.07)\tabularnewline
Men, 30\textendash 54 &
\textendash 0&31 (0.34) &
\textendash 1&50 (0.10) &
\textendash 0&47 (0.10) &
\textendash 0&82 (0.28) &
2&48 (0.32)\tabularnewline
Men, 55\textendash 74 &
1&51 (0.14) &
0&25 (0.03) &
0&25 (0.06) &
\textendash 0&01 (0.08) &
1&03 (0.12)\tabularnewline
Women, 16\textendash 29 &
1&07 (0.08) &
\textendash 0&10 (0.05) &
0&00 (0.04) &
\textendash 0&28 (0.08) &
1&45 (0.08)\tabularnewline
Women, 30\textendash 54 &
3&35 (0.18) &
\textendash 0&04 (0.04) &
0&53 (0.10) &
\textendash 0&89 (0.11) &
3&76 (0.16)\tabularnewline
Women, 55\textendash 74 &
0&77 (0.10) &
\textendash 0&98 (0.06) &
0&32 (0.05) &
\textendash 0&02 (0.05) &
1&46 (0.07)\tabularnewline
Total &
7&31 (0.72) &
\textendash 2&23 (0.19) &
0&48 (0.28) &
\textendash 2&14 (0.39) &
11&21 (0.56)\tabularnewline
\hline 
\end{tabular}
\par\end{centering}
{\scriptsize\textit{Notes}}{\scriptsize : All estimates are expressed
in log points multiplied by 100. Standard errors in parentheses are
clustered at the establishment level and also multiplied by 100. For
each period, the first column is the sum of the four components, and
the bottom row is the sum across worker types. The component labels
correspond to the four terms in equation \eqref{eq: formula}: \textquotedblleft demographic\textquotedblright{}
denotes demographic change, \textquotedblleft relative-share\textquotedblright{}
denotes relative-share change, \textquotedblleft relative-wage\textquotedblright{}
denotes relative-wage change, and \textquotedblleft cell-wage\textquotedblright{}
denotes cell-wage change. For the decomposition by worker type, $\check{w}_{g}$
in the demographic component is normalized by subtracting $\check{w}=(w_{t_{1}}+w_{t_{2}})/2$.}{\scriptsize\par}
\end{table}

Table \ref{tab: decomp_type1} also shows how the contribution of
each component is distributed across worker types. The demographic
component is negative in all three periods, but the contributions
differ across worker types. Figure \ref{fig: wage=000026share} helps
explain this pattern. Prime-age men have the highest wage levels throughout
the 1980\textendash 2024 period, but their employment share declines
substantially. The decline in their employment share substantially
lowers average log wages in every period. The decline in the employment
share of younger men raises average log wages in all three periods.
The decline in the employment share of younger women raises average
log wages before 2014, while the change in their employment share
after 2014 slightly lowers average log wages. At the same time, the
increase in the employment share of older men raises average log wages
in all three periods, while the increase in the employment share of
older women lowers average log wages in all three periods. The increase
in the employment share of prime-age women lowers average log wages
before 2014 but leaves them nearly unchanged after 2014. The negative
contributions nevertheless exceed the positive contributions in each
period, leaving the demographic component negative. The relative-share
component is negative for every worker type before 1996 and for most
worker types during the 1996\textendash 2014 period, but becomes positive
for several worker types after 2014. The relative-wage component is
positive for every worker type before 1996, with a particularly large
positive value for prime-age men, but is negative for most worker
types during the 1996\textendash 2014 period and for every worker
type after 2014. The cell-wage component shows a different pattern.
Before 1996, the cell-wage component is positive for every worker
type. During the 1996\textendash 2014 period, this component is negative
for men and non-negative for women. After 2014, it is positive again
for every worker type.

The figures in the appendix provide additional descriptive evidence
on differences in the levels and changes in hourly wages and employment
shares across job types for each worker type. Figures \ref{fig: wage=000026emp1_men}
and \ref{fig: wage=000026emp1_women} describe variation by employment
type and industry, while Figures \ref{fig: wage=000026emp2_men} and
\ref{fig: wage=000026emp2_women} describe variation by employment
type and establishment size. The figures illustrate substantial differences
in hourly wages and employment shares across employment types, industries,
and establishment sizes, as well as changes in these differences over
time. These patterns show substantial changes in employment shares
across the dimensions of job type over time, while the negative relative-share
component indicates that employment reallocation across job types
lowers average log wages before 2014. The figures also show that wage
changes after 2000 are uneven across job types. The negative relative-wage
component indicates that changes in relative wage structure across
job types lower average log wages after 1996. Table \ref{tab: decomp_type2}
in the appendix shows that the main results are robust to replacing
industry with region in the classification of job types.

\subsection{Employment reallocation and changes in relative wage structure\label{subsec: relative_emp=000026wage}}

The relative-share component captures employment reallocation across
job types, and the relative-wage component captures changes in relative
wage structure across job types. By themselves, however, these components
do not show how their contributions differ across employment type,
industry, and establishment size. A negative value of the relative-share
component indicates that, within worker types, on balance, employment
shifts toward job types with lower relative wages or away from job
types with higher relative wages. A negative value of the relative-wage
component indicates that, within worker types, on balance, relative
wages decline in job types with relatively high employment shares
or rise in job types with relatively low employment shares. The next
two subsections report the relative-share and relative-wage components
by employment type, industry, and establishment size.

\subsubsection{Sources of employment reallocation across job types\label{subsec: relative_emp}}

Table \ref{tab: relative_emp} reports the relative-share component
by job type. The first panel separates full-time and part-time jobs,
the second panel separates jobs by employment type and industry, and
the third panel separates jobs by employment type and establishment
size. During the 1980\textendash 1996 period, the relative-share component
is \textendash 1.14 log points for full-time jobs and \textendash 1.60
log points for part-time jobs. During the 1996\textendash 2014 period,
it is \textendash 0.99 and \textendash 1.94 log points, respectively.
During the 2014\textendash 2024 period, it is close to zero for full-time
jobs at \textendash 0.02 log points and positive for part-time jobs
at 0.50 log points. Thus, employment reallocation lowers average log
wages for both full-time and part-time job types before 2014, but
after 2014 it raises average log wages for part-time job types while
leaving average log wages nearly unchanged for full-time job types.

\begin{table}[h]
\caption{Relative-share component by job type\label{tab: relative_emp}}

\begin{centering}
\begin{tabular}{lr@{\extracolsep{0pt}.}lr@{\extracolsep{0pt}.}lr@{\extracolsep{0pt}.}l}
\hline 
 &
\multicolumn{2}{c}{1980\textendash 1996} &
\multicolumn{2}{c}{1996\textendash 2014} &
\multicolumn{2}{c}{2014\textendash 2024}\tabularnewline
\cline{2-7}
Total &
\textendash 2&74 (0.20) &
\textendash 2&94 (0.27) &
0&48 (0.28)\tabularnewline
\cline{2-7}
 &
\multicolumn{6}{c}{Panel A: Employment type}\tabularnewline
Full-time &
\textendash 1&14 (0.17) &
\textendash 0&99 (0.19) &
\textendash 0&02 (0.20)\tabularnewline
Part-time &
\textendash 1&60 (0.07) &
\textendash 1&94 (0.11) &
0&50 (0.11)\tabularnewline
\cline{2-7}
 &
\multicolumn{6}{c}{Panel B: Employment type $\times$ industry}\tabularnewline
Full-time $\times$ manufacturing &
\textendash 0&69 (0.14) &
\textendash 1&56 (0.17) &
\textendash 0&38 (0.17)\tabularnewline
Full-time $\times$ infrastructure &
\textendash 1&17 (0.12) &
\textendash 0&02 (0.19) &
\textendash 0&30 (0.20)\tabularnewline
Full-time $\times$ wholesale &
\textendash 0&10 (0.16) &
\textendash 0&86 (0.17) &
0&24 (0.18)\tabularnewline
Full-time $\times$ finance &
\textendash 0&61 (0.14) &
\textendash 0&58 (0.13) &
\textendash 0&11 (0.12)\tabularnewline
Full-time $\times$ services &
1&43 (0.13) &
2&01 (0.19) &
0&53 (0.19)\tabularnewline
Part-time $\times$ manufacturing &
\textendash 0&24 (0.04) &
0&43 (0.05) &
0&25 (0.03)\tabularnewline
Part-time $\times$ infrastructure &
\textendash 0&11 (0.01) &
\textendash 0&22 (0.02) &
0&05 (0.05)\tabularnewline
Part-time $\times$ wholesale &
\textendash 1&06 (0.06) &
\textendash 0&66 (0.08) &
0&29 (0.10)\tabularnewline
Part-time $\times$ finance &
\textendash 0&02 (0.00) &
\textendash 0&14 (0.01) &
0&04 (0.01)\tabularnewline
Part-time $\times$ services &
\textendash 0&16 (0.03) &
\textendash 1&36 (0.08) &
\textendash 0&13 (0.05)\tabularnewline
\cline{2-7}
 &
\multicolumn{6}{c}{Panel C: Employment type $\times$ establishment size}\tabularnewline
Full-time $\times$ medium+ &
\textendash 1&85 (0.20) &
\textendash 0&37 (0.24) &
\textendash 0&62 (0.25)\tabularnewline
Full-time $\times$ small &
0&71 (0.08) &
\textendash 0&62 (0.09) &
0&60 (0.09)\tabularnewline
Part-time $\times$ medium+ &
\textendash 0&51 (0.05) &
\textendash 0&67 (0.09) &
0&61 (0.08)\tabularnewline
Part-time $\times$ small &
\textendash 1&09 (0.05) &
\textendash 1&27 (0.09) &
\textendash 0&11 (0.10)\tabularnewline
\hline 
\end{tabular}
\par\end{centering}
{\scriptsize\textit{Notes}}{\scriptsize : All estimates are expressed
in log points multiplied by 100. Standard errors in parentheses are
clustered at the establishment level and also multiplied by 100. For
each period, each estimate is computed by summing the terms in the
relative-share component of equation \eqref{eq: formula} over the
job types included in that row. Each panel provides a separate decomposition
of the total relative-share component for the corresponding period.
Because the three panels decompose the same component in different
ways, estimates should be read within, not across, panels. The labels
\textquotedblleft medium+\textquotedblright{} and \textquotedblleft small\textquotedblright{}
denote establishments with 30 or more and fewer than 30 regularly
employed workers, respectively.}{\scriptsize\par}
\end{table}

The second panel shows that employment reallocation across job types
is not a simple shift from manufacturing to services. During the 1996\textendash 2014
period, the relative-share component has opposite signs between full-time
and part-time jobs in manufacturing and services. It is negative for
full-time jobs in manufacturing but positive for part-time jobs in
manufacturing, while the opposite pattern holds in services. Figures
\ref{fig: wage=000026emp1_men} and \ref{fig: wage=000026emp1_women}
illustrate these patterns. They show that wage levels differ substantially
between full-time and part-time jobs in services and manufacturing,
and that employment shares change differently across employment types
and industries for men and women. The negative relative-share component
therefore captures shifts across multi-dimensional job types, rather
than a simple movement from full-time to part-time employment or from
manufacturing to services.

The third panel shows that the relative-share component differs across
establishment sizes within employment type. During the 1996\textendash 2014
period, the relative-share component is negative for all four categories
defined by employment type and establishment size. The component is
relatively large and negative for part-time jobs in small establishments.
Figures \ref{fig: wage=000026emp2_men} and \ref{fig: wage=000026emp2_women}
show that full-time jobs in larger establishments generally have higher
wage levels, while part-time jobs and jobs in small establishments
generally have lower wage levels. They also show that employment shares
decline in full-time jobs in larger establishments for several male
worker types, while employment shares increase in part-time jobs and
jobs in small establishments, especially among women and older workers.
Because the relative-share component weights changes in relative employment
shares by relative log wages, shifts across establishment sizes can
affect average log wages even when employment type is held fixed.
After 2014, the relative-share component turns slightly positive,
but the pattern is not uniform across job types. Across employment
types, the relative-share component is positive for part-time jobs
and close to zero for full-time jobs. Across industries, it varies
in sign within both full-time and part-time employment. Across establishment
sizes, it is positive for full-time jobs in small establishments and
part-time jobs in larger establishments, but negative for full-time
jobs in larger establishments and part-time jobs in small establishments.
Overall, employment reallocation slightly raises average log wages
after 2014.

\subsubsection{Sources of changes in relative wage structure across job types\label{subsec: relative_wage}}

Table \ref{tab: relative_wage} reports the relative-wage component
by job type. As in Table \ref{tab: relative_emp}, the first panel
separates full-time and part-time jobs, the second panel separates
jobs by employment type and industry, and the third panel separates
jobs by employment type and establishment size. During the 1980\textendash 1996
period, the relative-wage component is positive for both full-time
and part-time jobs, at 5.92 and 5.64 log points, respectively. During
the 1996\textendash 2014 period, it turns negative at \textendash 0.40
log points for full-time jobs and \textendash 1.30 log points for
part-time jobs. During the 2014\textendash 2024 period, it remains
negative at \textendash 1.03 and \textendash 1.11 log points, respectively.
Thus, changes in relative wage structure raise average log wages for
both full-time and part-time job types before 1996 and lower average
log wages for both full-time and part-time job types after 1996.

\begin{table}[h]
\caption{Relative-wage component by job type\label{tab: relative_wage}}

\begin{centering}
\begin{tabular}{lr@{\extracolsep{0pt}.}lr@{\extracolsep{0pt}.}lr@{\extracolsep{0pt}.}l}
\hline 
 &
\multicolumn{2}{c}{1980\textendash 1996} &
\multicolumn{2}{c}{1996\textendash 2014} &
\multicolumn{2}{c}{2014\textendash 2024}\tabularnewline
\cline{2-7}
Total &
11&56 (1.10) &
\textendash 1&70 (0.48) &
\textendash 2&14 (0.39)\tabularnewline
\cline{2-7}
 &
\multicolumn{6}{c}{Panel A: Employment type}\tabularnewline
Full-time &
5&92 (0.56) &
\textendash 0&40 (0.29) &
\textendash 1&03 (0.27)\tabularnewline
Part-time &
5&64 (0.56) &
\textendash 1&30 (0.24) &
\textendash 1&11 (0.22)\tabularnewline
\cline{2-7}
 &
\multicolumn{6}{c}{Panel B: Employment type $\times$ industry}\tabularnewline
Full-time $\times$ manufacturing &
3&16 (0.31) &
0&05 (0.16) &
\textendash 0&42 (0.12)\tabularnewline
Full-time $\times$ infrastructure &
1&64 (0.20) &
0&00 (0.11) &
0&03 (0.10)\tabularnewline
Full-time $\times$ wholesale &
0&83 (0.11) &
0&03 (0.05) &
0&02 (0.03)\tabularnewline
Full-time $\times$ finance &
\textendash 0&45 (0.07) &
0&39 (0.07) &
0&40 (0.09)\tabularnewline
Full-time $\times$ services &
0&75 (0.10) &
\textendash 0&88 (0.13) &
\textendash 1&06 (0.17)\tabularnewline
Part-time $\times$ manufacturing &
\textendash 0&08 (0.21) &
\textendash 0&47 (0.14) &
\textendash 0&14 (0.12)\tabularnewline
Part-time $\times$ infrastructure &
0&10 (0.35) &
\textendash 0&18 (0.19) &
\textendash 0&17 (0.16)\tabularnewline
Part-time $\times$ wholesale &
\textendash 0&13 (0.32) &
\textendash 0&28 (0.14) &
0&01 (0.06)\tabularnewline
Part-time $\times$ finance &
3&67 (0.68) &
\textendash 0&92 (0.21) &
\textendash 0&99 (0.21)\tabularnewline
Part-time $\times$ services &
2&08 (0.50) &
0&55 (0.21) &
0&17 (0.08)\tabularnewline
\cline{2-7}
 &
\multicolumn{6}{c}{Panel C: Employment type $\times$ establishment size}\tabularnewline
Full-time $\times$ medium+ &
4&45 (0.43) &
\textendash 0&30 (0.25) &
\textendash 1&11 (0.23)\tabularnewline
Full-time $\times$ small &
1&47 (0.15) &
\textendash 0&10 (0.09) &
0&08 (0.09)\tabularnewline
Part-time $\times$ medium+ &
2&93 (0.58) &
\textendash 0&31 (0.27) &
\textendash 0&56 (0.20)\tabularnewline
Part-time $\times$ small &
2&71 (0.63) &
\textendash 0&99 (0.23) &
\textendash 0&55 (0.20)\tabularnewline
\hline 
\end{tabular}
\par\end{centering}
{\scriptsize\textit{Notes}}{\scriptsize : All estimates are expressed
in log points multiplied by 100. Standard errors in parentheses are
clustered at the establishment level and also multiplied by 100. For
each period, each estimate is computed by summing the terms in the
relative-wage component of equation \eqref{eq: formula} over the
job types included in that row. Each panel provides a separate decomposition
of the total relative-wage component for the corresponding period.
Because the three panels decompose the same component in different
ways, estimates should be read within, not across, panels. The labels
\textquotedblleft medium+\textquotedblright{} and \textquotedblleft small\textquotedblright{}
denote establishments with 30 or more and fewer than 30 regularly
employed workers, respectively.}{\scriptsize\par}
\end{table}

The second panel shows that changes in relative wage structure differ
across employment types and industries. During the 1996\textendash 2014
period, the relative-wage component is close to zero for full-time
jobs in manufacturing, infrastructure, and wholesale, positive for
full-time jobs in finance, and negative for full-time jobs in services.
For part-time jobs, it is negative in most industries but positive
in services. The component is relatively large and negative for part-time
jobs in finance and full-time jobs in services. The variation in signs
across industries within both full-time and part-time employment shows
that the contributions from changes in relative wage structure differ
across industries within employment type. After 2014, the relative-wage
component is negative for full-time jobs in manufacturing and services,
positive for full-time jobs in finance, and close to zero for full-time
jobs in infrastructure and wholesale. For part-time jobs, it is negative
in manufacturing, infrastructure, and finance, positive in services,
and close to zero in wholesale.

The third panel shows that the relative-wage component differs across
establishment sizes within employment type. During the 1996\textendash 2014
period, the relative-wage component is negative for all four categories
defined by employment type and establishment size. The component is
relatively large and negative for part-time jobs in small establishments.
After 2014, the relative-wage component remains negative for full-time
jobs in larger establishments and for part-time jobs in both establishment-size
categories, while it becomes slightly positive for full-time jobs
in small establishments.

\subsubsection{Summary}

Overall, Tables \ref{tab: relative_emp} and \ref{tab: relative_wage}
show that employment reallocation and changes in relative wage structure
are not limited to the distinction between full-time and part-time
jobs. During the 1996\textendash 2014 period, both employment reallocation
and changes in relative wage structure lower average log wages for
full-time job types as a whole and for part-time job types as a whole.
These changes also lower average log wages for full-time and part-time
jobs in both smaller and larger establishments. Across industries,
however, the direction differs. Within both full-time and part-time
employment, employment reallocation and changes in relative wage structure
each raise average log wages in some industries and lower average
log wages in others. These patterns show that the contributions from
employment reallocation and changes in relative wage structure differ
not only between full-time and part-time jobs but also across establishment
sizes and industries within employment type. These results complement
the literature on nonstandard employment and labor market dualization
by showing that the roles of employment reallocation and changes in
relative wage structure extend beyond differences across employment
types.

\subsection{Decomposition of changes in the gender wage gap\label{subsec: decomp_gap}}

In this section, I apply the decomposition framework to changes in
the gender wage gap. Let $f$ denote gender, where $f=M$ for men
and $f=F$ for women, and $w^{f}_{aht}$ denote average log wages
for workers of gender $f$ in age group $a$, job type $h$, and year
$t$. Average log wages can be written as the following weighted averages:
$w^{f}_{t}=\sum_{a}s^{f}_{at}w^{f}_{at}$ and $w^{f}_{at}=\sum_{h}s^{f}_{aht}w^{f}_{aht}$.
The employment shares are given by $s^{f}_{at}=n^{f}_{at}/\sum_{a}n^{f}_{at}$
and $s^{f}_{aht}=n^{f}_{aht}/\sum_{h}n^{f}_{aht}$, where $n^{f}_{aht}$
denotes weighted employment and $n^{f}_{at}=\sum_{h}n^{f}_{aht}$.
The gender wage gap is defined as $w^{M}_{t}-w^{F}_{t}$. For this
analysis, the decomposition framework is applied separately to men
and women. Worker types are defined by age group within each gender,
while job types are defined as in Section \ref{sec: data}. The change
in the gender wage gap is then decomposed into the differences in
the four components between men and women. A negative value of a component
indicates a narrowing of the gender wage gap, while a positive value
indicates a widening of the gap.

\begin{figure}[h]
\caption{Decomposition of changes in the gender wage gap\label{fig: decomp_gap}}

\begin{centering}
\includegraphics[scale=0.77]{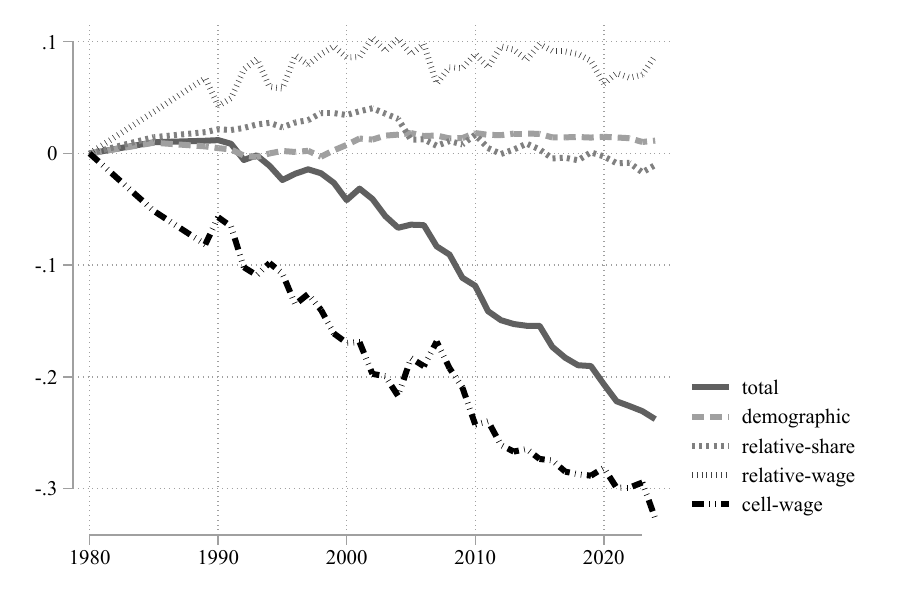}
\par\end{centering}
{\scriptsize\textit{Notes}}{\scriptsize : All series are normalized
to zero in 1980. For each year, the four components decompose the
change in the gender wage gap relative to 1980. The component labels
correspond to the four terms in equation \eqref{eq: formula}: \textquotedblleft demographic\textquotedblright{}
denotes demographic change, \textquotedblleft relative-share\textquotedblright{}
denotes relative-share change, \textquotedblleft relative-wage\textquotedblright{}
denotes relative-wage change, and \textquotedblleft cell-wage\textquotedblright{}
denotes cell-wage change.}{\scriptsize\par}
\end{figure}

Figure \ref{fig: decomp_gap} plots the change in the gender wage
gap relative to 1980 and the four decomposition components. The gender
wage gap narrows by 23.8 log points from 1980 to 2024. It narrows
gradually before the mid-1990s and more substantially thereafter.
These changes in the gender wage gap are not driven by a single component.
The cell-wage component declines substantially and consistently over
the period, apart from a temporary rise in the mid-2000s. At the same
time, the demographic component remains relatively small and positive,
while the relative-share component rises until around 2000 and declines
thereafter, becoming negative by 2024. The relative-wage component
rises until around 2000 and remains positive thereafter. The figure
therefore shows that the narrowing of the gender wage gap mainly involves
a decline in the cell-wage component, partly offset by the positive
relative-wage component.

Table \ref{tab: decomp_gap} reports the decomposition results by
age group for each period. Between 1980 and 1996, the gender wage
gap narrows by 1.8 log points. The cell-wage component is large and
negative. The relative-share and relative-wage components, however,
are positive, largely offsetting the negative cell-wage component.
The demographic component is also small and positive. Between 1996
and 2014, the gender wage gap narrows substantially by 13.6 log points.
This narrowing reflects not only a large negative cell-wage component,
but also negative relative-share and relative-wage components, while
the demographic component is positive. After 2014, the gender wage
gap narrows by a further 8.4 log points. The post-2014 narrowing is
mainly accounted for by the negative cell-wage component. The relative-share
component is also negative, while the relative-wage component is positive
and the demographic component is slightly negative. Across the three
periods, the cell-wage component is consistently negative, while the
demographic, relative-share, and relative-wage components vary in
sign. Notably, the gender wage gap narrows most during the 1996\textendash 2014
period, when average log wages fall rather than rise. The narrowing
of the gender wage gap during this period occurs alongside negative
unweighted mean wage growth for men, rather than positive growth for
both men and women.

\begin{table}[h]
\caption{Contributions by age group to changes in the gender wage gap\label{tab: decomp_gap}}

\begin{centering}
\begin{tabular}{lr@{\extracolsep{0pt}.}lr@{\extracolsep{0pt}.}lr@{\extracolsep{0pt}.}lr@{\extracolsep{0pt}.}lr@{\extracolsep{0pt}.}l}
\hline 
 &
\multicolumn{2}{c}{Total} &
\multicolumn{2}{c}{Demographic} &
\multicolumn{2}{c}{Relative-share} &
\multicolumn{2}{c}{Relative-wage} &
\multicolumn{2}{c}{Cell-wage}\tabularnewline
\cline{2-11}
 &
\multicolumn{10}{c}{1980\textendash 1996}\tabularnewline
16\textendash 29 &
\textendash 4&444 (0.206) &
0&782 (0.091) &
0&646 (0.095) &
\textendash 1&128 (0.593) &
\textendash 4&744 (0.632)\tabularnewline
30\textendash 54 &
1&030 (0.387) &
\textendash 0&721 (0.040) &
1&344 (0.272) &
8&320 (1.383) &
\textendash 7&913 (1.399)\tabularnewline
55\textendash 74 &
1&599 (0.136) &
0&070 (0.035) &
0&786 (0.073) &
1&591 (0.218) &
\textendash 0&848 (0.243)\tabularnewline
Total &
\textendash 1&814 (0.534) &
0&131 (0.127) &
2&776 (0.347) &
8&783 (1.638) &
\textendash 13&505 (1.674)\tabularnewline
\cline{2-11}
 &
\multicolumn{10}{c}{1996\textendash 2014}\tabularnewline
16\textendash 29 &
2&090 (0.173) &
2&404 (0.094) &
0&357 (0.117) &
0&554 (0.226) &
\textendash 1&226 (0.252)\tabularnewline
30\textendash 54 &
\textendash 12&999 (0.433) &
\textendash 0&292 (0.043) &
\textendash 2&136 (0.282) &
\textendash 0&445 (0.676) &
\textendash 10&126 (0.705)\tabularnewline
55\textendash 74 &
\textendash 2&700 (0.195) &
0&582 (0.043) &
\textendash 0&436 (0.112) &
\textendash 1&544 (0.180) &
\textendash 1&302 (0.218)\tabularnewline
Total &
\textendash 13&610 (0.621) &
2&693 (0.132) &
\textendash 2&214 (0.399) &
\textendash 1&435 (0.804) &
\textendash 12&654 (0.847)\tabularnewline
\cline{2-11}
 &
\multicolumn{10}{c}{2014\textendash 2024}\tabularnewline
16\textendash 29 &
\textendash 1&255 (0.179) &
\textendash 0&039 (0.089) &
\textendash 0&240 (0.087) &
0&377 (0.185) &
\textendash 1&353 (0.194)\tabularnewline
30\textendash 54 &
\textendash 5&801 (0.469) &
\textendash 0&396 (0.042) &
\textendash 2&034 (0.212) &
0&516 (0.537) &
\textendash 3&887 (0.540)\tabularnewline
55\textendash 74 &
\textendash 1&299 (0.251) &
0&279 (0.036) &
\textendash 0&248 (0.125) &
0&025 (0.180) &
\textendash 1&355 (0.217)\tabularnewline
Total &
\textendash 8&355 (0.728) &
\textendash 0&155 (0.120) &
\textendash 2&521 (0.343) &
0&917 (0.679) &
\textendash 6&595 (0.690)\tabularnewline
\hline 
\end{tabular}
\par\end{centering}
{\scriptsize\textit{Notes}}{\scriptsize : All estimates are expressed
in log points multiplied by 100. Standard errors in parentheses are
clustered at the establishment level and also multiplied by 100. For
each period, the first column is the sum of the four components, and
the bottom row is the sum across age groups. The component labels
correspond to the four terms in equation \eqref{eq: formula}: \textquotedblleft demographic\textquotedblright{}
denotes demographic change, \textquotedblleft relative-share\textquotedblright{}
denotes relative-share change, \textquotedblleft relative-wage\textquotedblright{}
denotes relative-wage change, and \textquotedblleft cell-wage\textquotedblright{}
denotes cell-wage change. For the decomposition by age group, $\check{w}^{f}_{a}$
in the demographic component is normalized by subtracting $\check{w}^{f}=(w^{f}_{t_{1}}+w^{f}_{t_{2}})/2$
within each gender.}{\scriptsize\par}
\end{table}

Table \ref{tab: decomp_gap} also shows how the contribution of each
component is distributed across age groups. The demographic component
differs in sign across age groups in all three periods; during the
1996\textendash 2014 period, the positive component for younger workers
accounts for most of the overall demographic component, while after
2014 offsetting signs leave the overall component close to zero. The
relative-share component is positive for all age groups before 1996.
After 1996, it is negative for most age groups, with the largest negative
values for prime-age workers in both periods. The relative-wage component
is particularly large and positive for prime-age workers before 1996,
turns negative for prime-age and older workers during the 1996\textendash 2014
period, and is positive for all age groups after 2014. The cell-wage
component is negative for every age group in all three periods and
is largest in absolute value for prime-age workers in each period.

Figures \ref{fig: wage=000026emp1_men}\textendash \ref{fig: wage=000026emp2_women}
show that part-time jobs generally pay less than full-time jobs and
account for a larger share of employment among women than among men.
These differences are relevant to the level of the gender wage gap.
However, the gender wage gap narrows, even as the share of part-time
employment increases more among women than among men within each age
group. Wage differences between men and women also narrow within several
job types, while wage growth in part-time jobs is not uniformly weaker
than in full-time jobs between 2000 and 2024. These patterns, together
with the decomposition results in Table \ref{tab: decomp_gap}, suggest
that the narrowing of the gender wage gap cannot be understood solely
in terms of the expansion of part-time employment among women.

\section{Conclusion\label{sec: conclusion}}

This paper reframes wage stagnation as the joint outcome of worker
composition, employment allocation, wage structure, and wage growth.
Applying the decomposition framework developed in this paper to repeated
cross-sectional data from a large-scale government survey in Japan
over the 45-year span from 1980 to 2024, I show how the four components
jointly account for the evolution of average log wages across periods
of wage growth, decline, and recovery. Before 1996, positive contributions
from unweighted mean wage growth and changes in relative wage structure
more than offset negative contributions from demographic change and
employment reallocation. During the 1996\textendash 2014 period, all
four components are negative. After 2014, the recovery is driven mainly
by renewed unweighted mean wage growth, while demographic change and
changes in relative wage structure remain negative and employment
reallocation turns slightly positive.

The results clarify the sources of Japan\textquoteright s wage stagnation
along four margins. First, wage stagnation is not simply a failure
of wages to rise on average across job types. Unweighted mean wage
growth raises average log wages before 1996 and after 2014, but lowers
average log wages during the 1996\textendash 2014 period. Second,
changes in worker composition lower average log wages in all three
periods. A major factor is the substantial decline in the employment
share of prime-age men, who have the highest wage levels. Third, employment
reallocation lowers average log wages before 2014. Fourth, changes
in relative wage structure lower average log wages after 1996. Employment
reallocation and changes in relative wage structure are not limited
to the distinction between full-time and part-time jobs and extend
across industries and establishment sizes within employment type.

The results also clarify the sources of the narrowing of Japan\textquoteright s
gender wage gap. Demographic change, employment reallocation, and
changes in relative wage structure vary in sign across periods, while
unweighted mean wage growth is the only force that narrows the gap
in all three periods. Notably, the gender wage gap narrows most during
the 1996\textendash 2014 period, when average log wages fall rather
than rise. Much of the narrowing after 1996 is concentrated among
workers aged 30 to 54.

These findings show that changes in both average log wages and the
gender wage gap reflect contributions from several distinct margins.
Explanations that focus on any single margin miss an important part
of the accounting.

\clearpage{}

\bibliographystyle{econ}
\bibliography{4C__Users_kenxy_Dropbox_yamada_06_payrise_paper_reference}

\clearpage{}

\appendix
\setcounter{equation}{0} \renewcommand{\theequation}{A\arabic{equation}}

\setcounter{table}{0} \renewcommand{\thetable}{A\arabic{table}}

\setcounter{figure}{0} \renewcommand{\thefigure}{A\arabic{figure}}

\section*{Appendix}

\section{Additional Descriptive Evidence}

\begin{figure}[h]
\caption{Male hourly wages and employment shares by employment type and industry\label{fig: wage=000026emp1_men}}

\begin{centering}
\subfloat[Men, 16\textendash 29]%
{
\centering{}\includegraphics[scale=0.58]{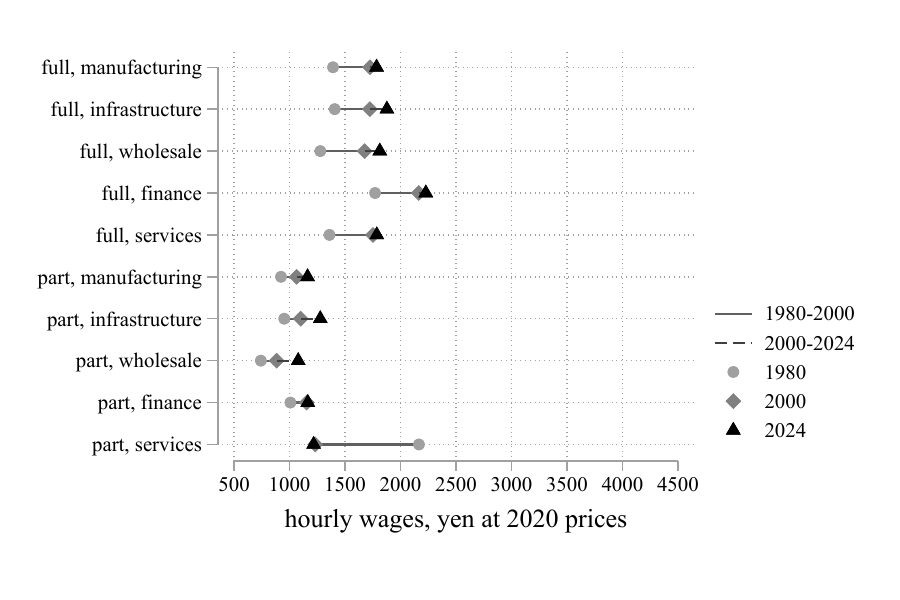}\includegraphics[scale=0.58]{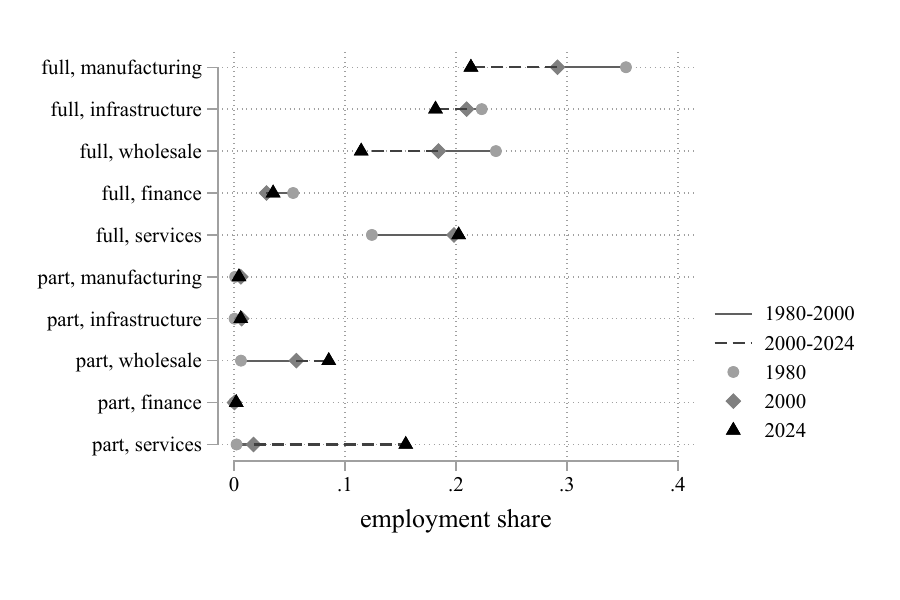}}
\par\end{centering}
\begin{centering}
\subfloat[Men, 30\textendash 54]%
{
\centering{}\includegraphics[scale=0.58]{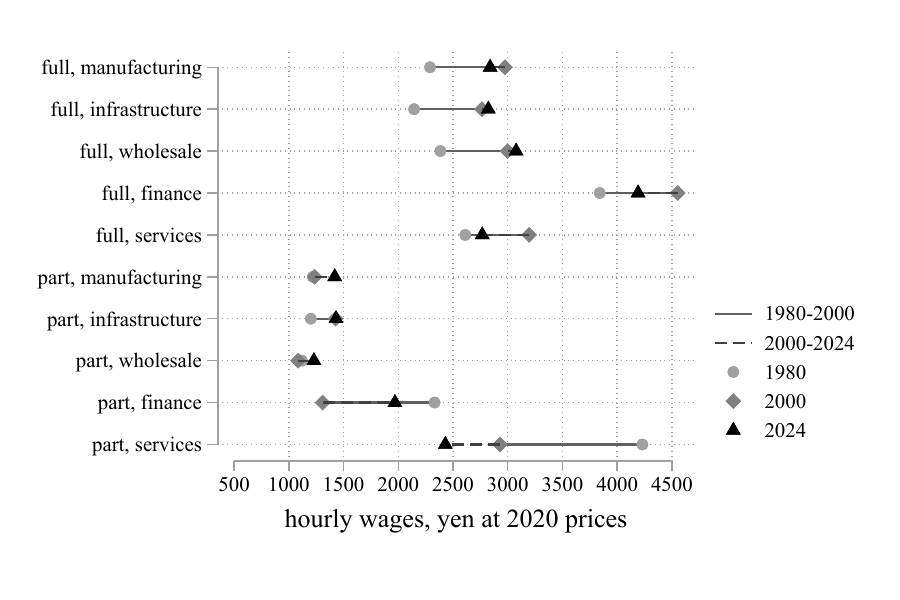}\includegraphics[scale=0.58]{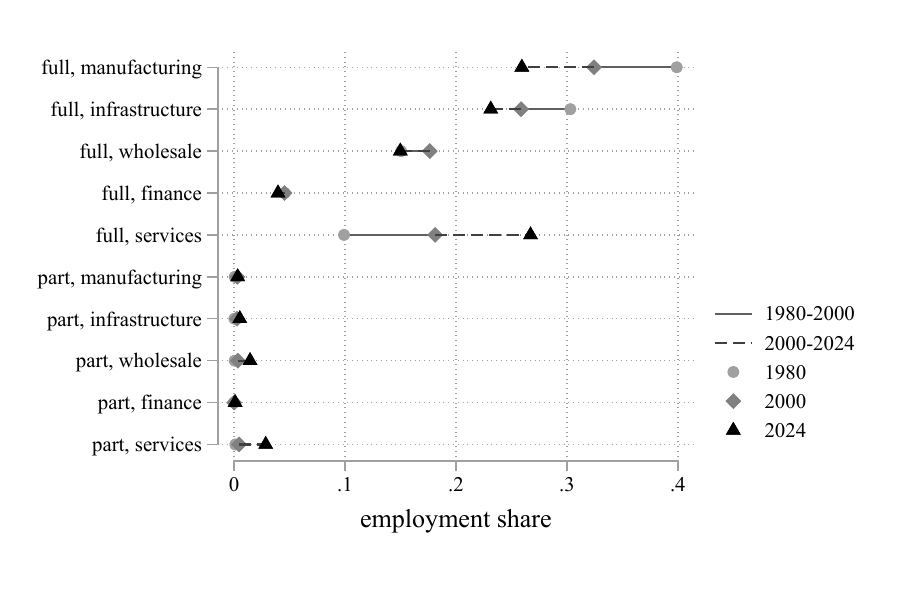}}
\par\end{centering}
\begin{centering}
\subfloat[Men, 55\textendash 74]%
{
\centering{}\includegraphics[scale=0.58]{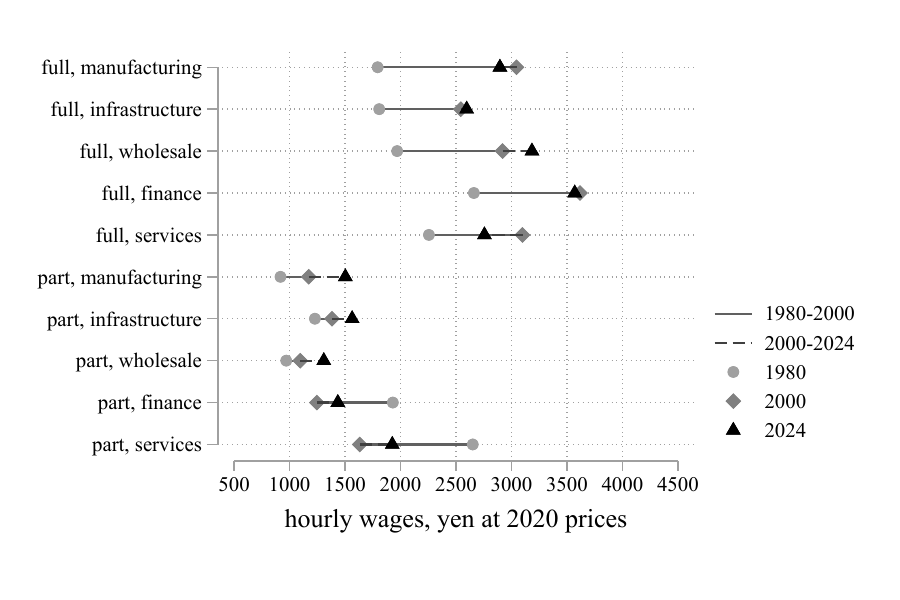}\includegraphics[scale=0.58]{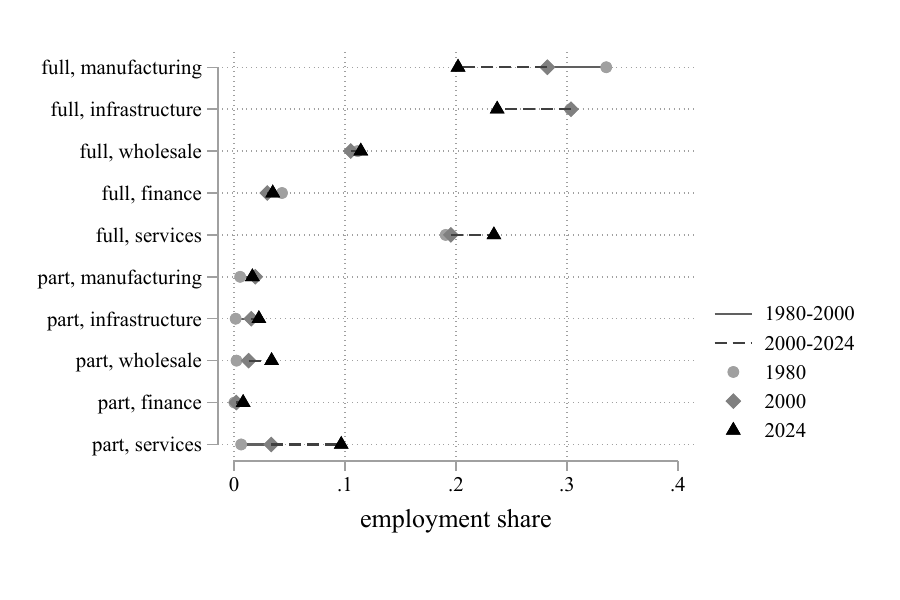}}
\par\end{centering}
{\scriptsize\textit{Notes}}{\scriptsize : Hourly wages are measured
in yen and deflated to 2020 prices. Employment shares are measured
as shares of employment within each worker type, so that they sum
to one across the job types shown in each panel. The line segments
connect 1980 to 2000 and 2000 to 2024, respectively. Job types shown
here are grouped by employment type and industry; establishment size
is not distinguished. The labels \textquotedblleft full\textquotedblright{}
and \textquotedblleft part\textquotedblright{} denote full-time and
part-time jobs, respectively.}{\scriptsize\par}
\end{figure}

\begin{figure}[h]
\caption{Female hourly wages and employment shares by employment type and industry\label{fig: wage=000026emp1_women}}

\begin{centering}
\subfloat[Women, 16\textendash 29]%
{
\centering{}\includegraphics[scale=0.58]{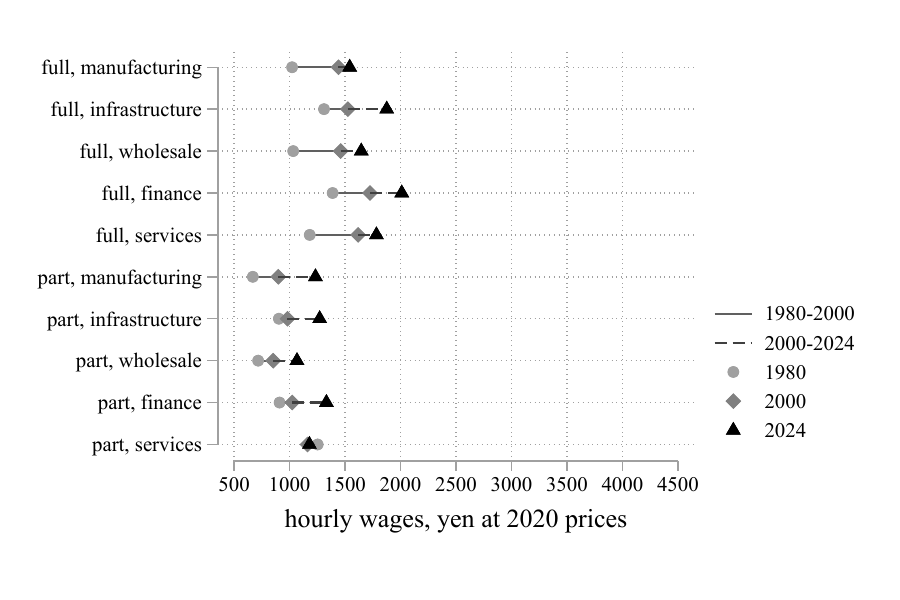}\includegraphics[scale=0.58]{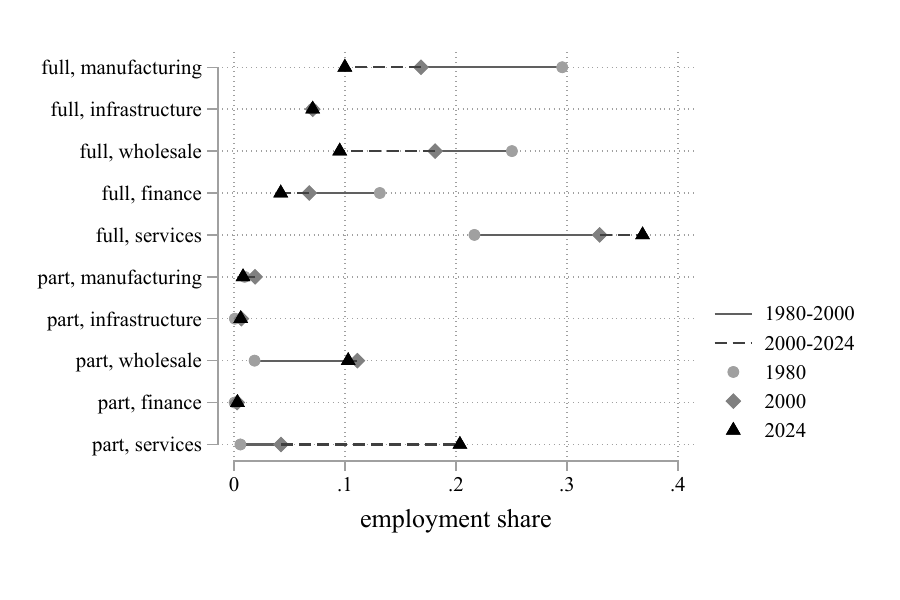}}
\par\end{centering}
\begin{centering}
\subfloat[Women, 30\textendash 54]%
{
\centering{}\includegraphics[scale=0.58]{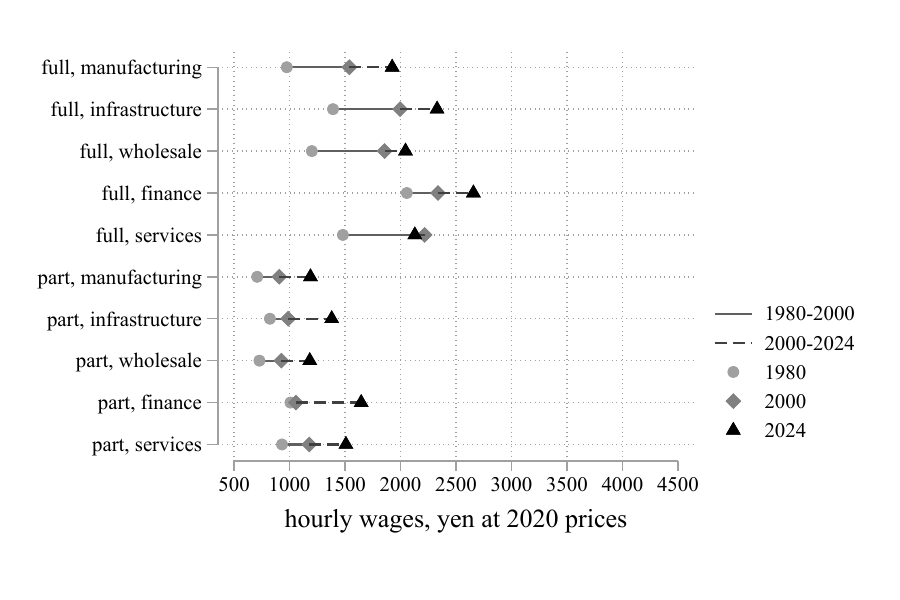}\includegraphics[scale=0.58]{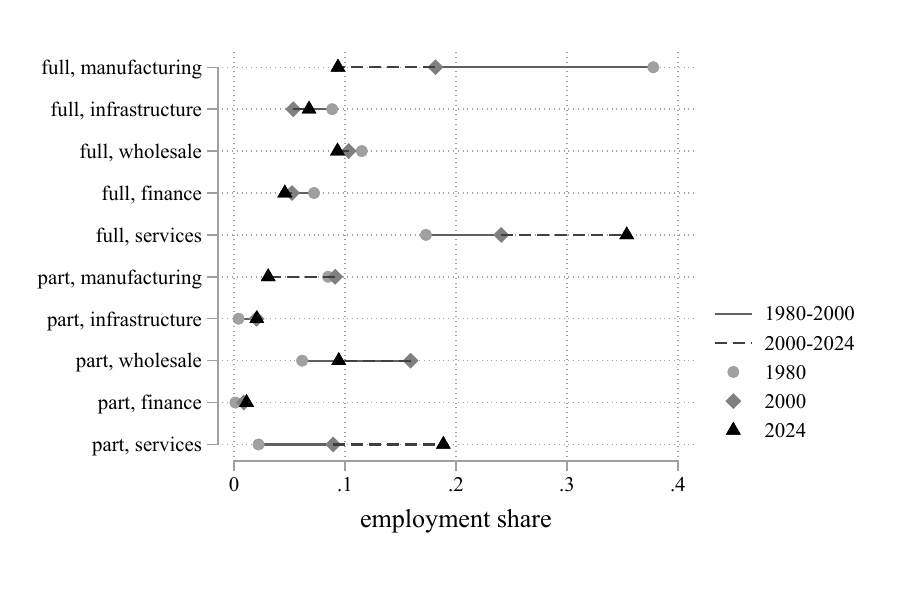}}
\par\end{centering}
\begin{centering}
\subfloat[Women, 55\textendash 74]%
{
\centering{}\includegraphics[scale=0.58]{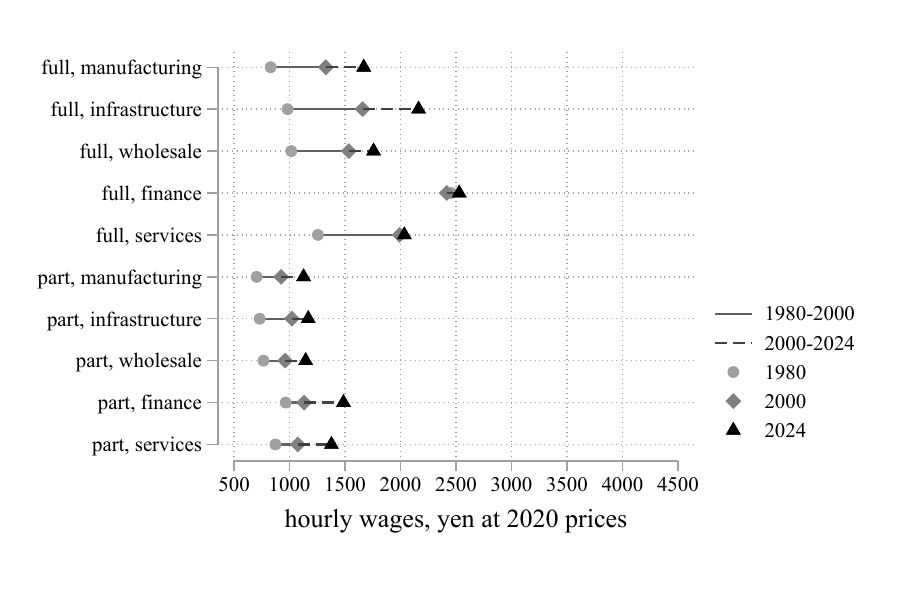}\includegraphics[scale=0.58]{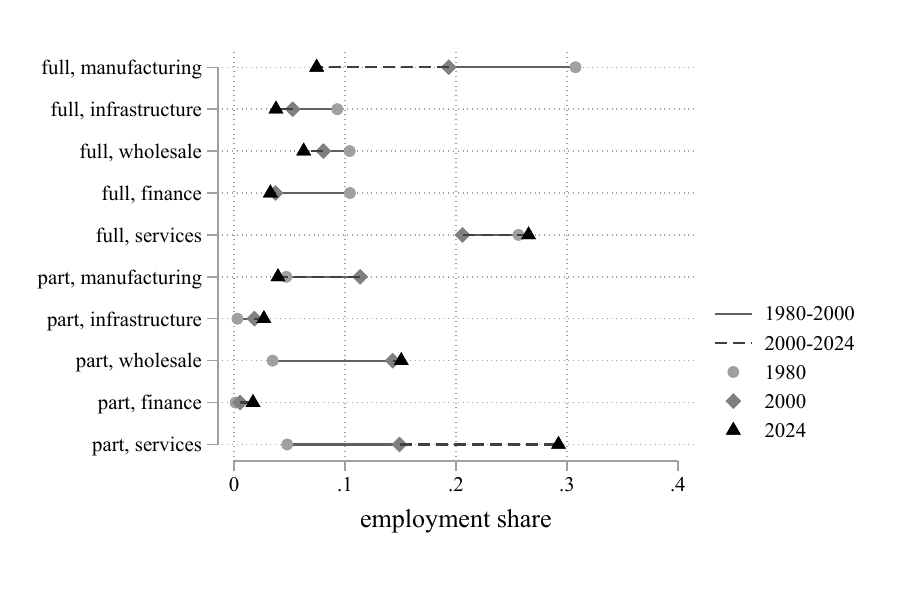}}
\par\end{centering}
{\scriptsize\textit{Notes}}{\scriptsize : Hourly wages are measured
in yen and deflated to 2020 prices. Employment shares are measured
as shares of employment within each worker type, so that they sum
to one across the job types shown in each panel. The line segments
connect 1980 to 2000 and 2000 to 2024, respectively. Job types shown
here are grouped by employment type and industry; establishment size
is not distinguished. The labels \textquotedblleft full\textquotedblright{}
and \textquotedblleft part\textquotedblright{} denote full-time and
part-time jobs, respectively.}{\scriptsize\par}
\end{figure}

\begin{figure}[h]
\caption{Male hourly wages and employment shares by employment type and establishment
size\label{fig: wage=000026emp2_men}}

\begin{centering}
\subfloat[Men, 16\textendash 29]%
{
\centering{}\includegraphics[scale=0.58]{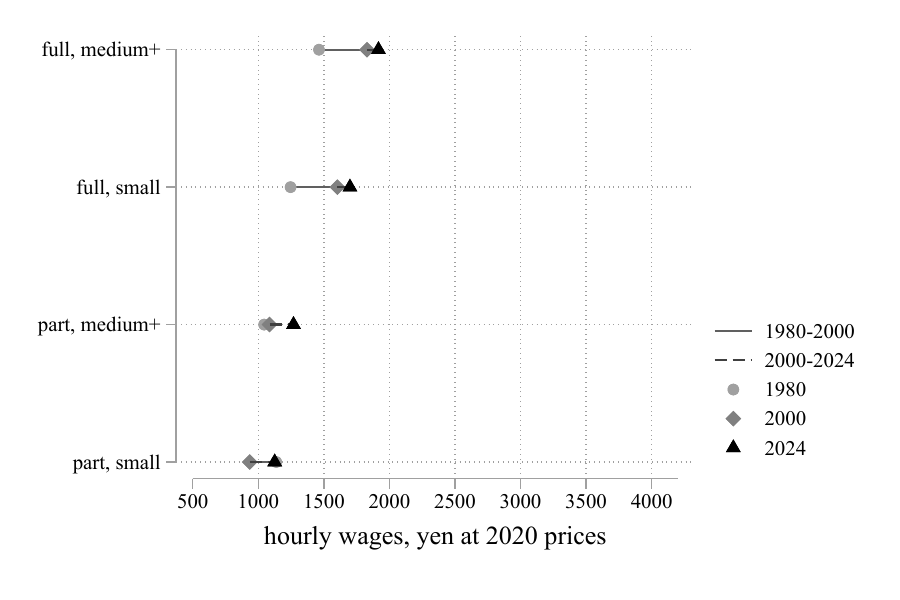}\includegraphics[scale=0.58]{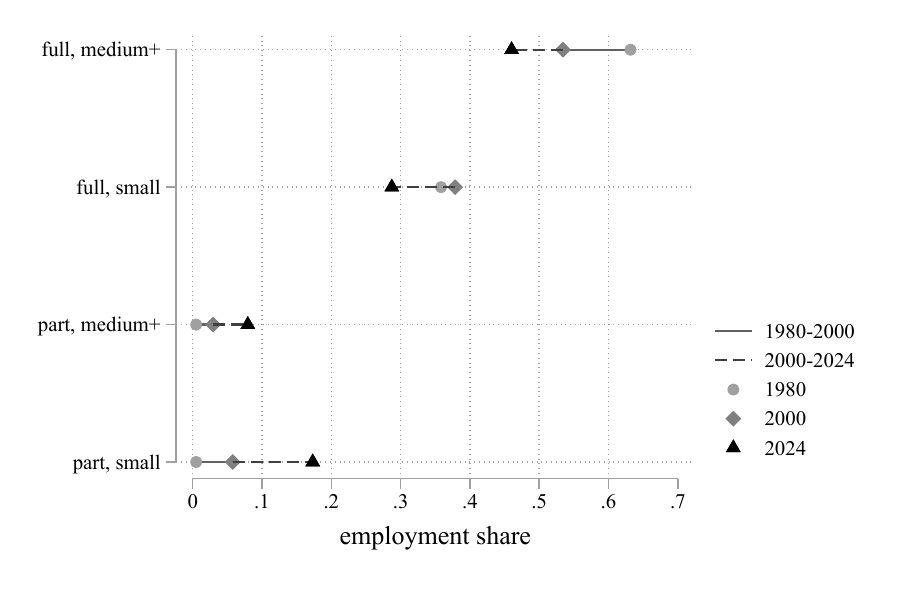}}
\par\end{centering}
\begin{centering}
\subfloat[Men, 30\textendash 54]%
{
\centering{}\includegraphics[scale=0.58]{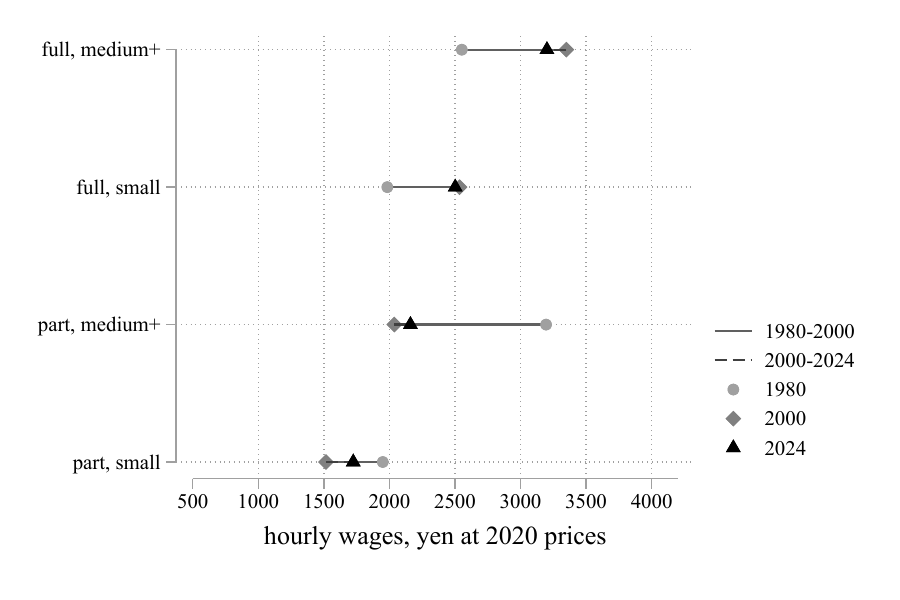}\includegraphics[scale=0.58]{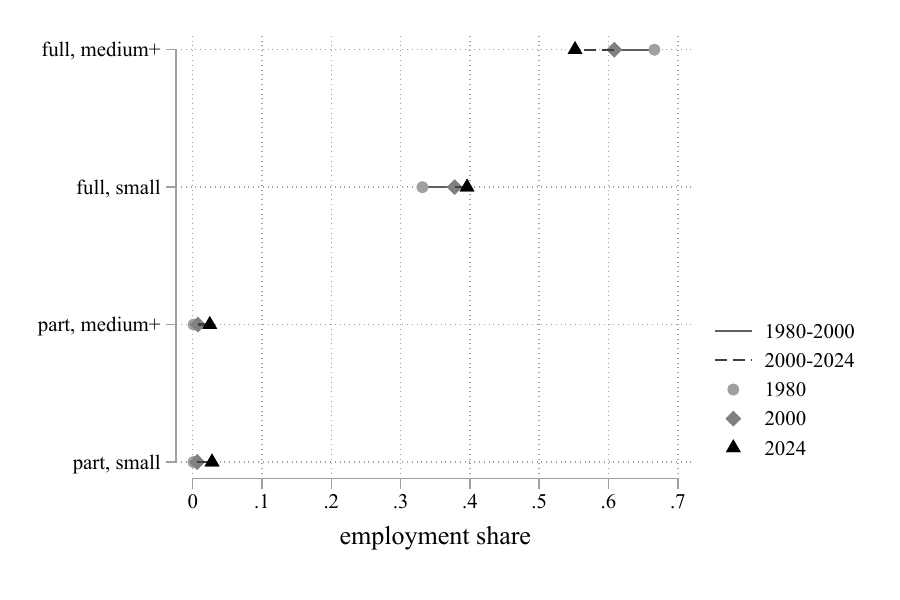}}
\par\end{centering}
\begin{centering}
\subfloat[Men, 55\textendash 74]%
{
\centering{}\includegraphics[scale=0.58]{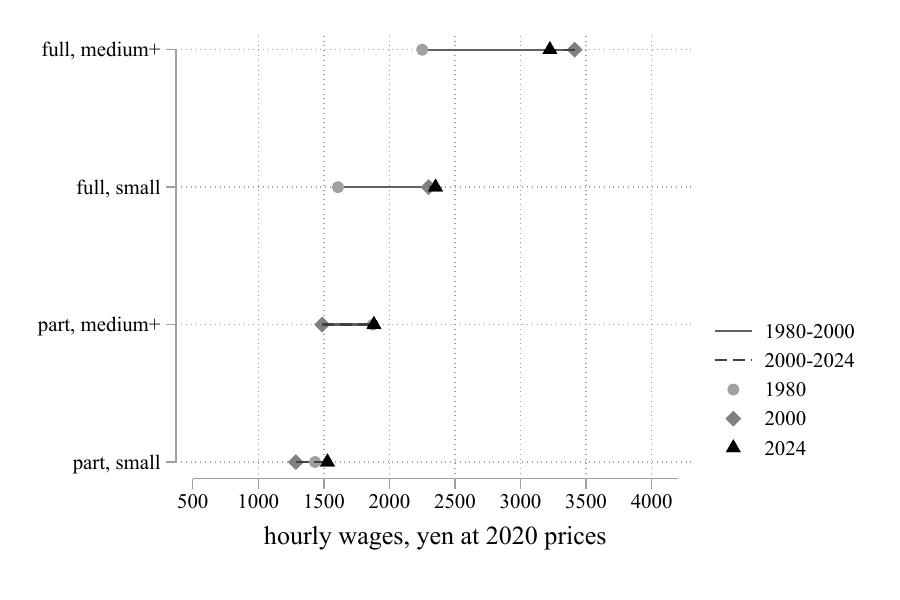}\includegraphics[scale=0.58]{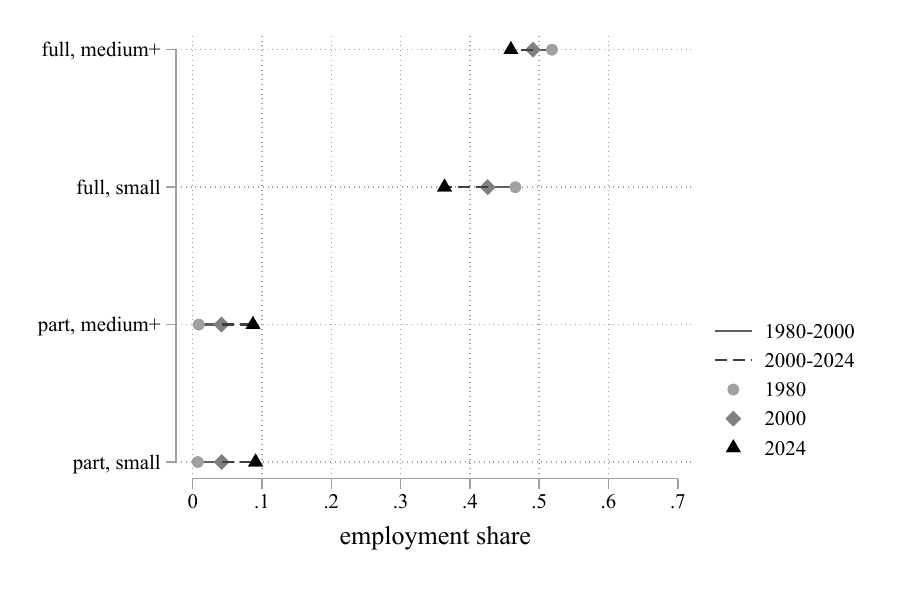}}
\par\end{centering}
{\scriptsize\textit{Notes}}{\scriptsize : Hourly wages are measured
in yen and deflated to 2020 prices. Employment shares are measured
as shares of employment within each worker type, so that they sum
to one across the job types shown in each panel. The line segments
connect 1980 to 2000 and 2000 to 2024, respectively. Job types shown
here are grouped by employment type and establishment size; industry
is not distinguished. The labels \textquotedblleft full\textquotedblright{}
and \textquotedblleft part\textquotedblright{} denote full-time and
part-time jobs, respectively. The labels \textquotedblleft medium+\textquotedblright{}
and \textquotedblleft small\textquotedblright{} denote establishments
with 30 or more and fewer than 30 regularly employed workers, respectively.}{\scriptsize\par}
\end{figure}

\begin{figure}[h]
\caption{Female hourly wages and employment shares by employment type and establishment
size\label{fig: wage=000026emp2_women}}

\begin{centering}
\subfloat[Women, 16\textendash 29]%
{
\centering{}\includegraphics[scale=0.58]{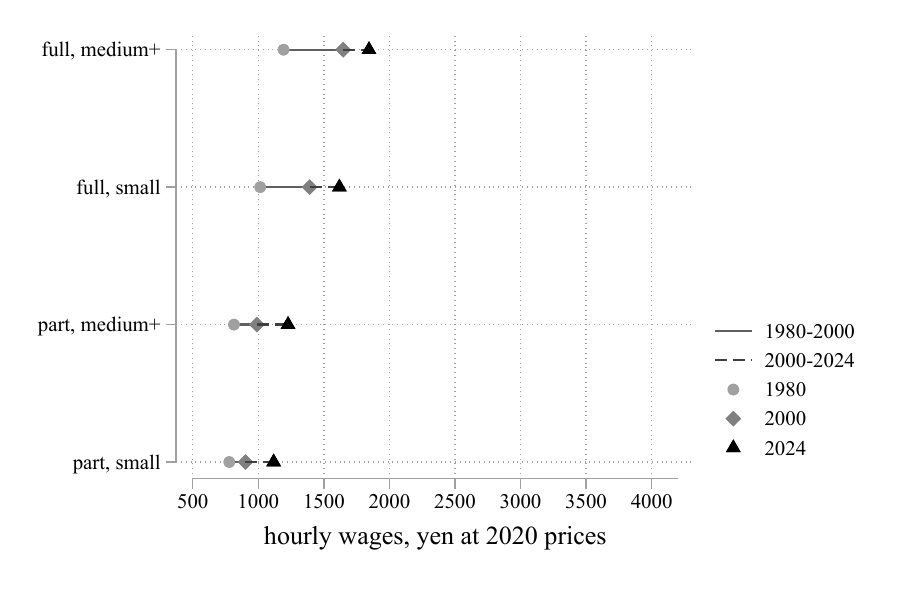}\includegraphics[scale=0.58]{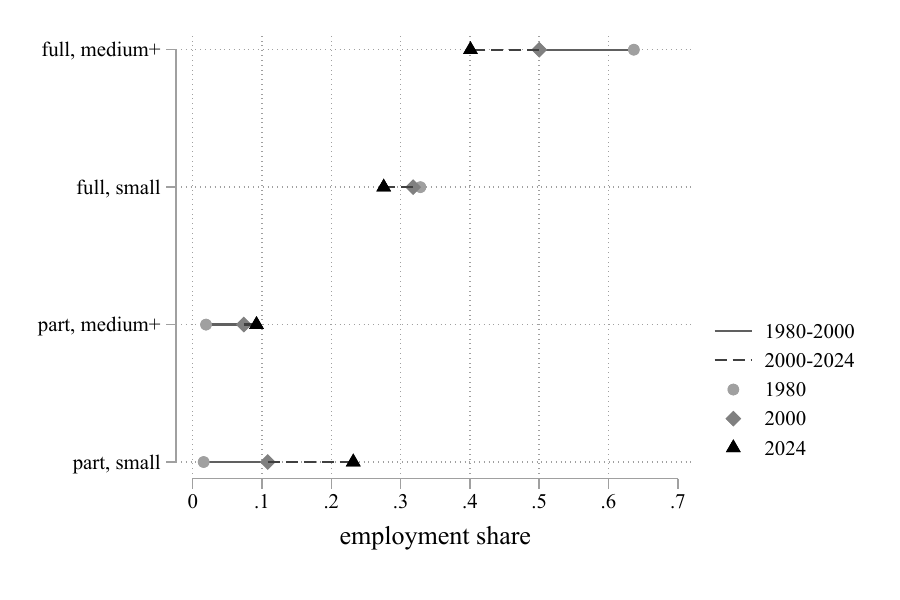}}
\par\end{centering}
\begin{centering}
\subfloat[Women, 30\textendash 54]%
{
\centering{}\includegraphics[scale=0.58]{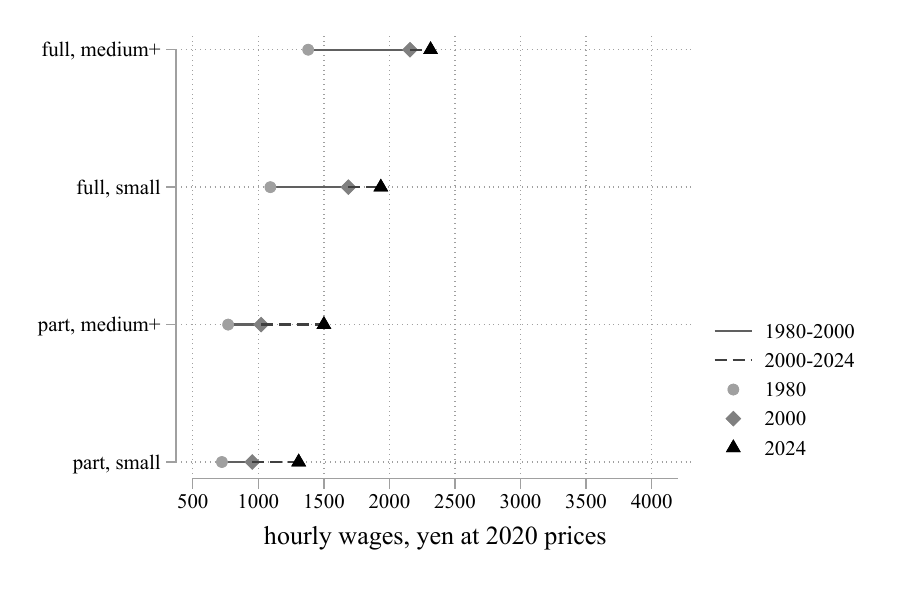}\includegraphics[scale=0.58]{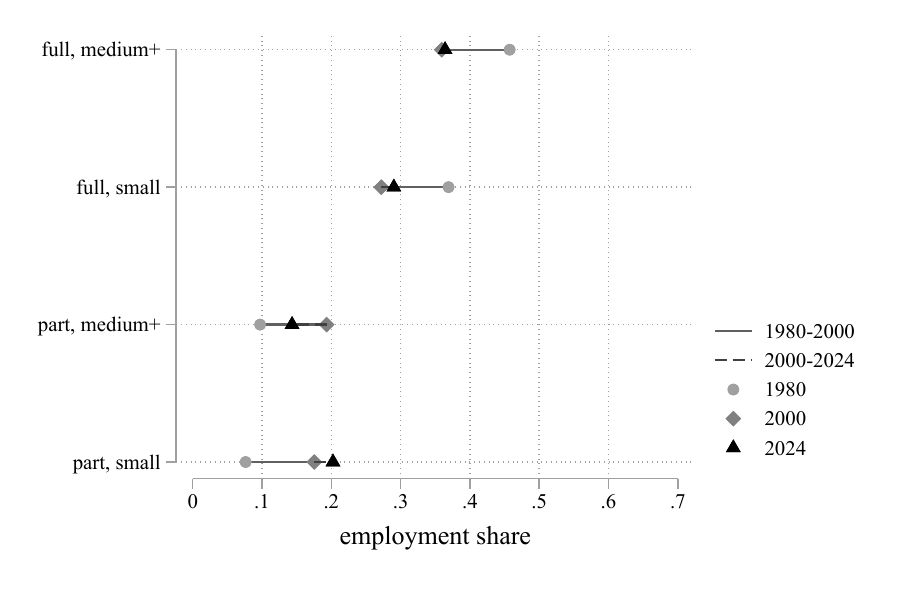}}
\par\end{centering}
\begin{centering}
\subfloat[Women, 55\textendash 74]%
{
\centering{}\includegraphics[scale=0.58]{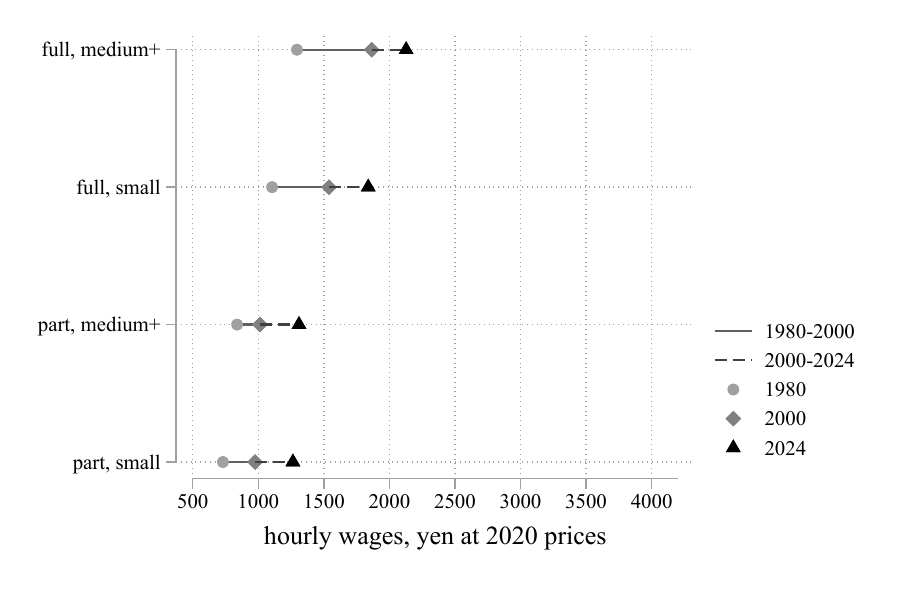}\includegraphics[scale=0.58]{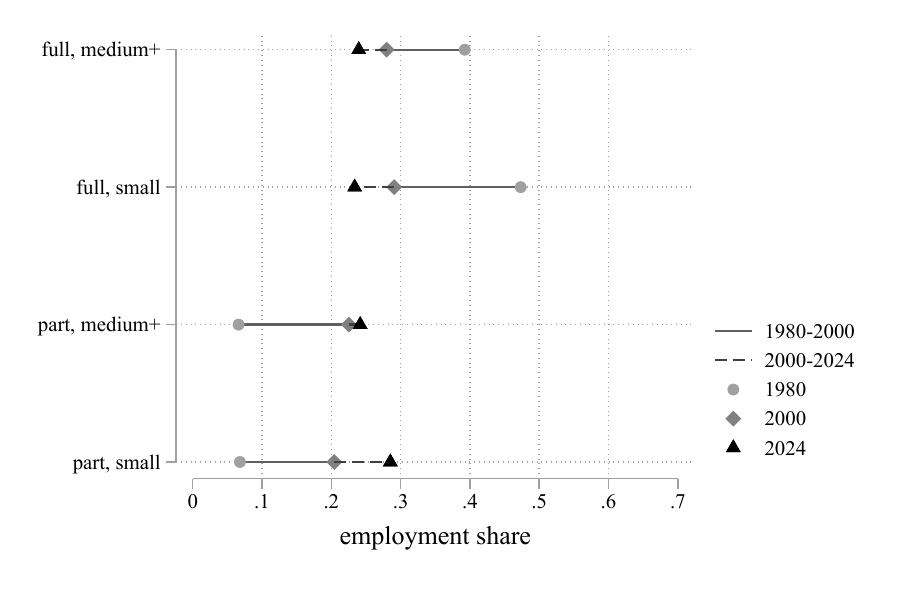}}
\par\end{centering}
{\scriptsize\textit{Notes}}{\scriptsize : Hourly wages are measured
in yen and deflated to 2020 prices. Employment shares are measured
as shares of employment within each worker type, so that they sum
to one across the job types shown in each panel. The line segments
connect 1980 to 2000 and 2000 to 2024, respectively. Job types shown
here are grouped by employment type and establishment size; industry
is not distinguished. The labels \textquotedblleft full\textquotedblright{}
and \textquotedblleft part\textquotedblright{} denote full-time and
part-time jobs, respectively. The labels \textquotedblleft medium+\textquotedblright{}
and \textquotedblleft small\textquotedblright{} denote establishments
with 30 or more and fewer than 30 regularly employed workers, respectively.}{\scriptsize\par}
\end{figure}

This section documents differences in the levels and changes in hourly
wages and employment shares across job types for each worker type.
Figures \ref{fig: wage=000026emp1_men} and \ref{fig: wage=000026emp1_women}
show differences in hourly wages and employment shares by employment
type and industry separately for men and women. Figures \ref{fig: wage=000026emp2_men}
and \ref{fig: wage=000026emp2_women} present the corresponding evidence
by employment type and establishment size. As in Figure \ref{fig: wage=000026share},
the line segments connect 1980 to 2000 and 2000 to 2024, respectively.
The figures show that many job types experience wage increases before
2000 but weaker wage growth or wage declines after 2000, while employment
shares often change more steadily over time. Several patterns are
visible. First, full-time job types generally have higher wage levels
than part-time job types, but there is also substantial variation
across industries within full-time and part-time employment. Second,
wage levels and employment shares also differ by establishment size
within employment type, with larger establishments generally associated
with higher full-time wages. Third, employment is not distributed
evenly across job types. Full-time jobs in several industries account
for large shares of male employment, while employment among women
is more concentrated in services and part-time jobs. Fourth, the employment
shares within each worker type change over time, indicating that changes
in employment shares are observed not only between full-time and part-time
employment but also across industries and establishment sizes within
employment type. These patterns support the distinction between worker
types and job types and provide a basis for examining employment reallocation
and changes in relative wage structure across job types within worker
types.

\section{Robustness Checks}

This section examines whether the main results are robust to alternative
classifications of job types and an alternative measure of wages.

\subsection{Alternative classifications of job types}

\begin{figure}[h]
\caption{Decomposition of changes in average log real hourly wages when job
types are defined by employment type, establishment size, and region\label{fig: decomp_region}}

\begin{centering}
\includegraphics[scale=0.77]{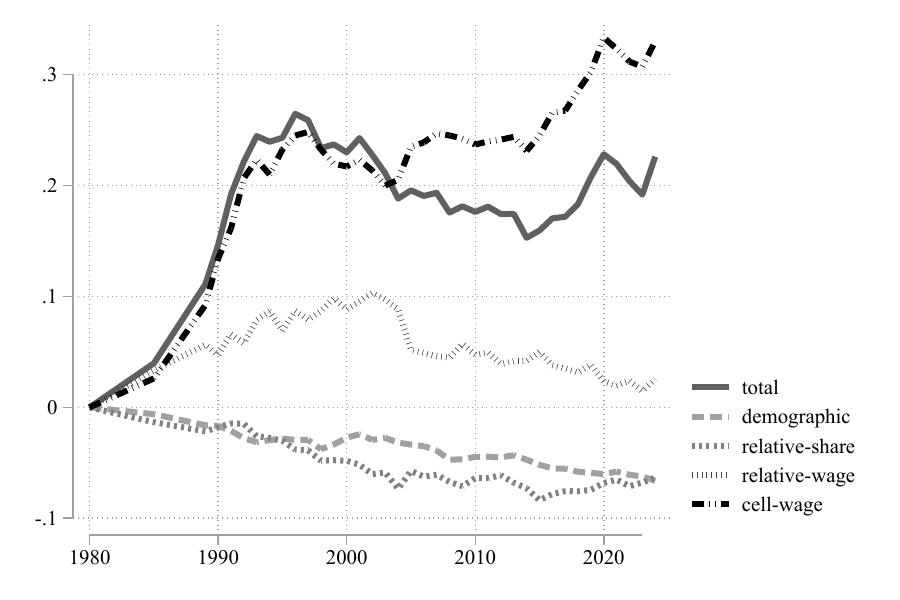}
\par\end{centering}
{\scriptsize\textit{Notes}}{\scriptsize : All series are normalized
to zero in 1980. For each year, the four components decompose the
change in average log wages relative to 1980. The component labels
correspond to the four terms in equation \eqref{eq: formula}: \textquotedblleft demographic\textquotedblright{}
denotes demographic change, \textquotedblleft relative-share\textquotedblright{}
denotes relative-share change, \textquotedblleft relative-wage\textquotedblright{}
denotes relative-wage change, and \textquotedblleft cell-wage\textquotedblright{}
denotes cell-wage change.}{\scriptsize\par}
\end{figure}

I first examine whether the main results depend on the choice of job
attributes in the classification of job types. Figure \ref{fig: decomp_region}
presents the decomposition over time using an alternative classification
of job types, in which industry is replaced by region as one dimension
of job type. Table \ref{tab: decomp_type2} reports the contributions
by worker type for the three periods under the same alternative classification.
In the baseline specification, job types are defined by employment
type, establishment size, and industry. In the alternative specification,
industry is replaced by region, while worker types remain defined
by sex and age group. Regions are defined as ten broad regional blocks:
Hokkaido and Tohoku, Kita-Kanto, Minami-Kanto, Hokuriku, Tosan, Tokai,
Kinki, Chugoku, Shikoku, and Kyushu and Okinawa.

Figure \ref{fig: decomp_region} shows that the main patterns of the
decomposition over time are similar to those in Figure \ref{fig: decomp_main}.
The demographic component does not depend on the classification of
job types and therefore remains identical, whereas the other three
components change in magnitude when industry is replaced by region.
The relative-share component remains negative over most of the 1980\textendash 2024
period, while the relative-wage component rises until around 2000
and declines thereafter. The cell-wage component rises strongly before
the mid-1990s and again after the mid-2010s.

Table \ref{tab: decomp_type2} shows that the relative-share, relative-wage,
and cell-wage components retain the same signs as in Table \ref{tab: decomp_type1}
in each of the three periods. Before 1996, unweighted mean wage growth
and changes in relative wage structure raise average log wages, while
demographic change and employment reallocation lower them. During
the 1996\textendash 2014 period, all four components are negative.
After 2014, unweighted mean wage growth raises average log wages,
and employment reallocation no longer lowers them, while demographic
change and changes in relative wage structure continue to lower average
log wages. These results indicate that the main patterns of the decomposition
are robust to replacing industry with region in the classification
of job types.

\begin{table}[h]
\caption{Contributions by worker type to changes in average log real hourly
wages when job types are defined by employment type, establishment
size, and region\label{tab: decomp_type2}}

\begin{centering}
\begin{tabular}{lr@{\extracolsep{0pt}.}lr@{\extracolsep{0pt}.}lr@{\extracolsep{0pt}.}lr@{\extracolsep{0pt}.}lr@{\extracolsep{0pt}.}l}
\hline 
 &
\multicolumn{2}{c}{Total} &
\multicolumn{2}{c}{Demographic} &
\multicolumn{2}{c}{Relative-share} &
\multicolumn{2}{c}{Relative-wage} &
\multicolumn{2}{c}{Cell-wage}\tabularnewline
\cline{2-11}
 &
\multicolumn{10}{c}{1980\textendash 1996}\tabularnewline
Men, 16\textendash 29 &
4&41 (0.08) &
0&37 (0.03) &
\textendash 0&50 (0.04) &
1&62 (0.27) &
2&93 (0.29)\tabularnewline
Men, 30\textendash 54 &
9&84 (0.24) &
\textendash 1&98 (0.10) &
\textendash 0&90 (0.10) &
4&99 (0.88) &
7&74 (0.90)\tabularnewline
Men, 55\textendash 74 &
3&26 (0.07) &
0&46 (0.02) &
\textendash 0&11 (0.03) &
0&91 (0.12) &
2&00 (0.13)\tabularnewline
Women, 16\textendash 29 &
4&20 (0.09) &
0&21 (0.06) &
\textendash 0&55 (0.04) &
0&55 (0.10) &
3&98 (0.11)\tabularnewline
Women, 30\textendash 54 &
4&86 (0.16) &
\textendash 0&89 (0.08) &
\textendash 1&46 (0.08) &
0&56 (0.07) &
6&66 (0.09)\tabularnewline
Women, 55\textendash 74 &
\textendash 0&10 (0.06) &
\textendash 1&10 (0.05) &
\textendash 0&29 (0.02) &
0&08 (0.03) &
1&22 (0.03)\tabularnewline
Total &
26&48 (0.48) &
\textendash 2&94 (0.21) &
\textendash 3&82 (0.21) &
8&71 (1.03) &
24&52 (1.06)\tabularnewline
\cline{2-11}
 &
\multicolumn{10}{c}{1996\textendash 2014}\tabularnewline
Men, 16\textendash 29 &
\textendash 0&63 (0.08) &
1&03 (0.03) &
\textendash 1&14 (0.05) &
\textendash 0&39 (0.08) &
\textendash 0&13 (0.09)\tabularnewline
Men, 30\textendash 54 &
\textendash 7&31 (0.29) &
\textendash 1&36 (0.10) &
\textendash 0&51 (0.13) &
\textendash 1&42 (0.50) &
\textendash 4&02 (0.52)\tabularnewline
Men, 55\textendash 74 &
\textendash 1&42 (0.10) &
0&54 (0.03) &
\textendash 0&88 (0.06) &
\textendash 0&69 (0.09) &
\textendash 0&39 (0.09)\tabularnewline
Women, 16\textendash 29 &
0&28 (0.08) &
1&24 (0.05) &
\textendash 1&10 (0.05) &
\textendash 0&16 (0.05) &
0&30 (0.06)\tabularnewline
Women, 30\textendash 54 &
\textendash 0&36 (0.17) &
\textendash 1&48 (0.07) &
\textendash 0&31 (0.11) &
\textendash 0&30 (0.07) &
1&72 (0.10)\tabularnewline
Women, 55\textendash 74 &
\textendash 1&74 (0.09) &
\textendash 1&73 (0.06) &
\textendash 0&68 (0.04) &
\textendash 0&12 (0.03) &
0&78 (0.04)\tabularnewline
Total &
\textendash 11&18 (0.59) &
\textendash 1&76 (0.21) &
\textendash 4&61 (0.30) &
\textendash 3&08 (0.57) &
\textendash 1&73 (0.64)\tabularnewline
\cline{2-11}
 &
\multicolumn{10}{c}{2014\textendash 2024}\tabularnewline
Men, 16\textendash 29 &
0&91 (0.08) &
0&14 (0.03) &
\textendash 0&10 (0.05) &
\textendash 0&06 (0.05) &
0&93 (0.05)\tabularnewline
Men, 30\textendash 54 &
\textendash 0&31 (0.34) &
\textendash 1&50 (0.10) &
\textendash 0&21 (0.13) &
\textendash 0&69 (0.37) &
2&09 (0.37)\tabularnewline
Men, 55\textendash 74 &
1&51 (0.14) &
0&25 (0.03) &
0&36 (0.07) &
0&04 (0.10) &
0&86 (0.09)\tabularnewline
Women, 16\textendash 29 &
1&07 (0.08) &
\textendash 0&10 (0.05) &
0&11 (0.05) &
0&01 (0.04) &
1&06 (0.05)\tabularnewline
Women, 30\textendash 54 &
3&35 (0.18) &
\textendash 0&04 (0.04) &
0&67 (0.11) &
\textendash 0&29 (0.09) &
3&02 (0.11)\tabularnewline
Women, 55\textendash 74 &
0&77 (0.10) &
\textendash 0&98 (0.06) &
0&32 (0.04) &
0&05 (0.04) &
1&39 (0.06)\tabularnewline
Total &
7&31 (0.72) &
\textendash 2&23 (0.19) &
1&16 (0.33) &
\textendash 0&95 (0.51) &
9&34 (0.54)\tabularnewline
\hline 
\end{tabular}
\par\end{centering}
{\scriptsize\textit{Notes}}{\scriptsize : All estimates are expressed
in log points multiplied by 100. Standard errors in parentheses are
clustered at the establishment level and also multiplied by 100. For
each period, the first column is the sum of the four components, and
the bottom row is the sum across worker types. The component labels
correspond to the four terms in equation \eqref{eq: formula}: \textquotedblleft demographic\textquotedblright{}
denotes demographic change, \textquotedblleft relative-share\textquotedblright{}
denotes relative-share change, \textquotedblleft relative-wage\textquotedblright{}
denotes relative-wage change, and \textquotedblleft cell-wage\textquotedblright{}
denotes cell-wage change. Job types are defined by employment type,
establishment size, and region. For the decomposition by worker type,
$\check{w}_{g}$ in the demographic component is normalized by subtracting
$\check{w}=(w_{t_{1}}+w_{t_{2}})/2$.}{\scriptsize\par}
\end{table}

\begin{figure}[h]
\caption{Decomposition of changes in average log real hourly wages when job
types are defined by employment type and establishment size\label{fig: decomp_noindustry}}

\begin{centering}
\includegraphics[scale=0.77]{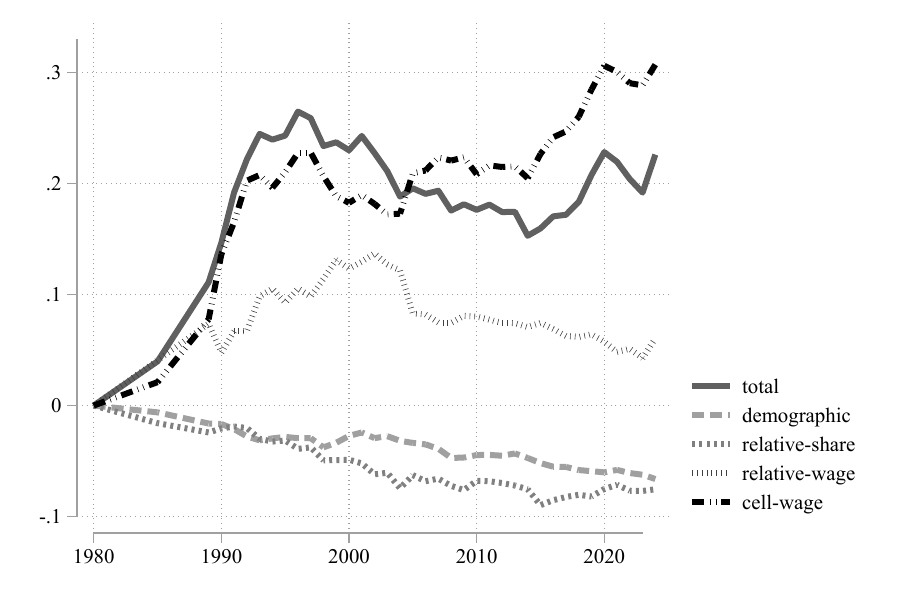}
\par\end{centering}
{\scriptsize\textit{Notes}}{\scriptsize : All series are normalized
to zero in 1980. For each year, the four components decompose the
change in average log wages relative to 1980. The component labels
correspond to the four terms in equation \eqref{eq: formula}: \textquotedblleft demographic\textquotedblright{}
denotes demographic change, \textquotedblleft relative-share\textquotedblright{}
denotes relative-share change, \textquotedblleft relative-wage\textquotedblright{}
denotes relative-wage change, and \textquotedblleft cell-wage\textquotedblright{}
denotes cell-wage change.}{\scriptsize\par}
\end{figure}

\begin{figure}[h]
\caption{Decomposition of changes in average log real hourly wages when job
types are defined by employment type and industry\label{fig: decomp_nosize}}

\begin{centering}
\includegraphics[scale=0.77]{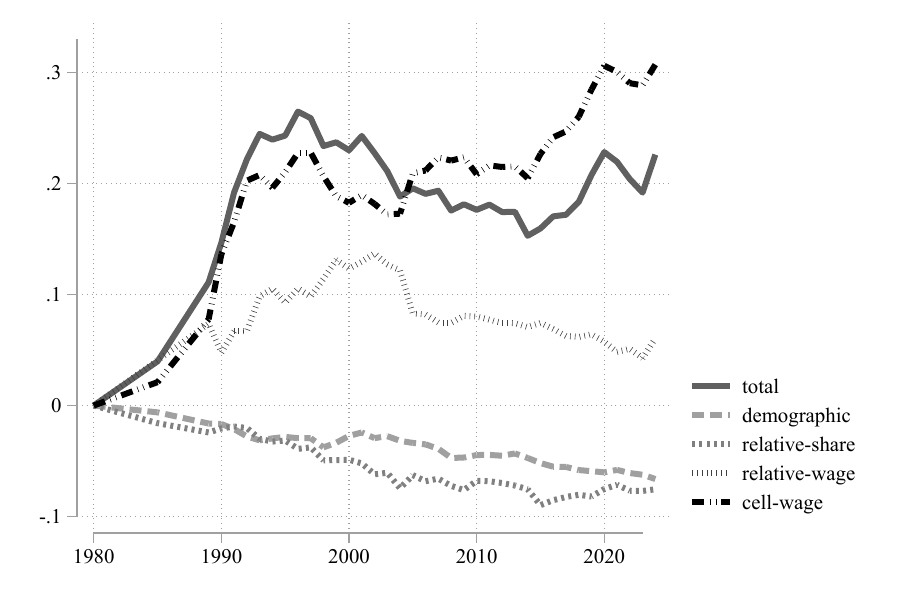}
\par\end{centering}
{\scriptsize\textit{Notes}}{\scriptsize : All series are normalized
to zero in 1980. For each year, the four components decompose the
change in average log wages relative to 1980. The component labels
correspond to the four terms in equation \eqref{eq: formula}: \textquotedblleft demographic\textquotedblright{}
denotes demographic change, \textquotedblleft relative-share\textquotedblright{}
denotes relative-share change, \textquotedblleft relative-wage\textquotedblright{}
denotes relative-wage change, and \textquotedblleft cell-wage\textquotedblright{}
denotes cell-wage change.}{\scriptsize\par}
\end{figure}

I next examine whether the main results depend on the granularity
of the classification of job types. The cell-wage component is the
largest in absolute value among the four components in each of the
three periods in Table \ref{tab: decomp_type1}, but its magnitude
may vary with how finely job types are classified. In the baseline
specification, job types are classified by three job attributes: employment
type, establishment size, and industry. In the alternative specifications,
job types are classified more coarsely by two of the three job attributes.
Figure \ref{fig: decomp_noindustry} presents the decomposition when
job types are defined by employment type and establishment size. Figure
\ref{fig: decomp_nosize} presents the decomposition when job types
are defined by employment type and industry. Both figures show that
the main patterns of the decomposition over time are similar to those
in Figure \ref{fig: decomp_main}. These results indicate that the
main patterns of the decomposition are robust to the two coarser classifications
of job types.

\subsection{Alternative measure of wages}

\begin{figure}[h]
\caption{Decomposition of changes in average log real hourly wages, excluding
bonuses\label{fig: decomposition_nobonus}}

\begin{centering}
\includegraphics[scale=0.77]{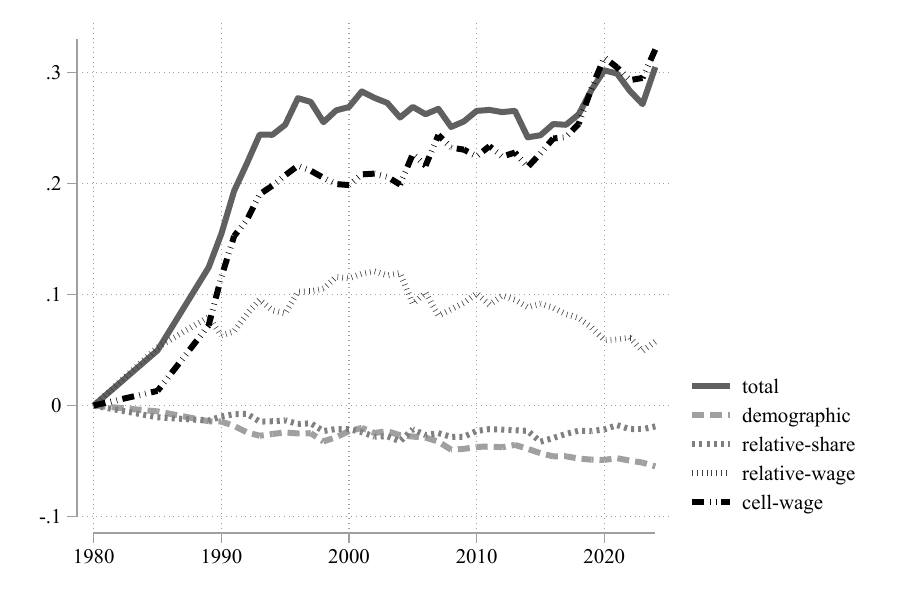}
\par\end{centering}
{\scriptsize\textit{Notes}}{\scriptsize : All series are normalized
to zero in 1980. For each year, the four components decompose the
change in average log wages relative to 1980. The component labels
correspond to the four terms in equation \eqref{eq: formula}: \textquotedblleft demographic\textquotedblright{}
denotes demographic change, \textquotedblleft relative-share\textquotedblright{}
denotes relative-share change, \textquotedblleft relative-wage\textquotedblright{}
denotes relative-wage change, and \textquotedblleft cell-wage\textquotedblright{}
denotes cell-wage change.}{\scriptsize\par}
\end{figure}

I finally examine whether the main results depend on the inclusion
of annual bonuses in hourly wages. Figure \ref{fig: decomposition_nobonus}
presents the decomposition using hourly wages that exclude annual
bonuses. As in Figure \ref{fig: decomp_main}, average log wages rise
rapidly before the mid-1990s. Thereafter, however, the decline is
smaller when annual bonuses are excluded. Average log wages remain
relatively stable through much of the 2000s and early 2010s.

The demographic and relative-share components remain negative over
much of the 1980\textendash 2024 period, while the relative-wage component
rises until around 2000 and declines thereafter. The main difference
is in the cell-wage component. Its decline after the mid-1990s is
smaller when annual bonuses are excluded, accounting for much of the
difference in the path of average log wages between Figures \ref{fig: decomp_main}
and \ref{fig: decomposition_nobonus}. The main patterns of the other
three components remain essentially unchanged.
\end{document}